\documentclass{ptephy_v1}

\usepackage{graphics} 

\usepackage{dcolumn}
\usepackage{bm}
\usepackage{amsfonts}
\usepackage{latexsym}
\usepackage{amssymb}
\usepackage{verbatim}
\usepackage{amsthm}
\usepackage{amsmath}
\usepackage{mathrsfs}
\usepackage{wrapft}

\newcommand{\FF}{{{\mathbb F}}}

\newcommand{\RF}{{{\mathbb R}}}

\newcommand{\ScrF}{{{\mathscr F}}}
\newcommand{\ScrG}{{{\mathscr G}}}
\newcommand{\ScrH}{{{\mathscr H}}}

\newcommand{\ScrM}{{{\mathscr M}}}

\newcommand{\ScrQ}{{{\mathscr Q}}}

\newcommand{\ScrT}{{{\mathscr T}}}

\newcommand{\ScrX}{{{\mathscr X}}}
\newcommand{\ScrY}{{{\mathscr Y}}}

\newtheorem{theorem}{Theorem}[section]

\newtheorem{conjecture}{Conjecture}[section]
\newtheorem{proposal}{Proposal}[section]

\begin{document}


\numberwithin{equation}{section}

\title{
  Gauge-invariant perturbation theory on the Schwarzschild
  background spacetime Part II: \\
  --- Even-mode perturbations ---
}
\author{
  Kouji Nakamura
  \footnote{E-mail address: dr.kouji.nakamura@gmail.com}
}
\address{
  Gravitational-Wave Science Project,
  National Astronomical Observatory of Japan,\\
  2-21-1, Osawa, Mitaka, Tokyo 181-8588, Japan
}
\date{\today}
\begin{abstract}
  This is the Part II paper of our series of papers on a
  gauge-invariant perturbation theory on the Schwarzschild background
  spacetime.
  After reviewing our general framework of the gauge-invariant
  perturbation theory and the proposal on the gauge-invariant
  treatments for $l=0,1$ mode perturbations on the Schwarzschild
  background spacetime in the Part I paper [K.~Nakamura,
  arXiv:2110.13508 [gr-qc]], we examine the linearized Einstein
  equations for even-mode perturbations.
  We discuss the strategy to solve the linearized Einstein equations
  for these even-mode perturbations including $l=0,1$ modes.
  Furthermore, we explicitly derive the $l=0,1$ mode solutions to the
  linearized Einstein equations in both the vacuum and the non-vacuum
  cases.
  We show that the solutions for $l=0$ mode perturbations includes the
  additional Schwarzschild mass parameter perturbation, which is
  physically reasonable.
  Then, we conclude that our proposal of the resolution of the
  $l=0,1$-mode problem is physically reasonable due to the realization
  of the additional Schwarzschild mass parameter perturbation and the
  Kerr parameter perturbation in the Part I paper.
\end{abstract}

\maketitle


\section{Introduction}
\label{sec:introduction}


Gravitational-wave observations are now carrying out through the
ground-based
detectors~\cite{LIGO-home-page,Virgo-home-page,KAGRA-home-page,LIGO-INDIA-home-page}.
Furthermore, the projects of future ground-based gravitational-wave
detectors~\cite{ET-home-page,CosmicExplorer-home-page} are also
progressing to achieve more sensitive detectors.
In addition to these ground-based detectors, some projects of space
gravitational-wave antenna are also
progressing~\cite{LISA-home-page,DECIGO-PTEP-2021,TianQin-PTEP-2021,Taiji-PTEP-2021}.
Among them, the Extreme-Mass-Ratio-Inspiral (EMRI), which is a source
of gravitational waves from the motion of a stellar mass object around
a supermassive black hole, is a promising target of the Laser
Interferometer Space Antenna~\cite{LISA-home-page}.
To describe the gravitational wave from EMRIs, black hole
perturbations are used~\cite{L.Barack-A.Pound-2019}.
Furthermore, the sophistication of higher-order black hole
perturbation theories is required to support these gravitational-wave
physics as a precise science.
The motivation of this paper is in such theoretical sophistications of
black hole perturbation theories toward higher-order perturbations for
wide physical situations.


Although realistic black holes have their angular momentum and we have
to consider the perturbation theory of a Kerr black hole for direct
applications to the EMRI, we may say that further sophistications are
possible even in perturbation theories on the Schwarzschild background
spacetime.
From the pioneering works by Regge and
Wheeler~\cite{T.Regge-J.A.Wheeler-1957} and
Zerilli~\cite{F.Zerilli-1970-PRL,F.Zerilli-1970-PRD,H.Nakano-2019},
there have been many studies on the perturbations in the Schwarzschild
background
spacetime~\cite{H.Nakano-2019,V.Moncrief-1974a,V.Moncrief-1974b,C.T.Cunningham-R.H.Price-V.Moncrief-1978,Chandrasekhar-1983,Gerlach_Sengupta-1979a,Gerlach_Sengupta-1979b,Gerlach_Sengupta-1979c,Gerlach_Sengupta-1980,T.Nakamura-K.Oohara-Y.Kojima-1987,Gundlach-Martine-Garcia-2000,Gundlach-Martine-Garcia-2001,A.Nagar-L.Rezzolla-2005-2006,K.Martel-E.Poisson-2005}.
In these works, perturbations on the Schwarzschild spacetime are
decomposed through the spherical harmonics $Y_{lm}$ because of the
spherical symmetry of the background spacetime, and $l=0$ and $l=1$
modes should be separately treated.
Furthermore, ``{\it gauge-invariant}'' treatments for $l=0$ and $l=1$
even-modes were unknown.


Owing to this situation, in the previous
papers~\cite{K.Nakamura-2021a,K.Nakamura-2021c}, we proposed the
strategy of the gauge-invariant treatments of these $l=0,1$ mode
perturbations, which is declared as
Proposal~\ref{proposal:treatment-proposal-on-pert-on-spherical-BG}
in Sec.~\ref{sec:review-of-general-framework-GI-perturbation-theroy}
of this paper below.
One of important premises of our gauge-invariant perturbations is the
distinction of the first-kind gauge and the second-kind gauge.
The first-kind gauge is the choice of the coordinate system on the
single manifold and we often use this first-kind gauge when we predict
or interpret the measurement results of experiments and observation.
On the other hand, the second-kind gauge is the choice of the
point-identifications between the points on the physical spacetime
$\ScrM_{\epsilon}$ and the background spacetime $\ScrM$.
This second-kind gauge have nothing to do with our physical spacetime
$\ScrM$.
The proposal in the Part I paper~\cite{K.Nakamura-2021c} is a part of
our developments of the general formulation of a higher-order
gauge-invariant perturbation theory on a generic background spacetime
toward unambiguous sophisticated nonlinear general-relativistic
perturbation theories~\cite{K.Nakamura-2003,K.Nakamura-2005,K.Nakamura-2011,K.Nakamura-IJMPD-2012,K.Nakamura-2013,K.Nakamura-2014}.
This general formulation of the higher-order gauge-invariant
perturbation theory was applied to cosmological perturbations~\cite{K.Nakamura-2006,K.Nakamura-2007,K.Nakamura-2008,K.Nakamura-2009a,K.Nakamura-2009b,K.Nakamura-2010,A.J.Christopherson-K.A.Malik-D.R.Matravers-K.Nakamura-2011,K.Nakamura-2020}.
Even in cosmological perturbation theories, the same problem as the
above $l=0,1$-mode problem exists as gauge-invariant treatments of
homogeneous modes of perturbations.
In this sense, we can expect that the proposal in the previous
paper~\cite{K.Nakamura-2021c} will be a clue to the same problem in
gauge-invariant perturbation theory on the generic background
spacetime.


In addition to the proposal of the gauge-invariant treatments of
$l=0,1$-mode perturbations on the Schwarzschild background spacetime,
in the previous Part I paper, we also derived the linearized Einstein
equations in a gauge-invariant manner following
Proposal~\ref{proposal:treatment-proposal-on-pert-on-spherical-BG}.
From the parity of perturbations, we can classify the perturbations on
the spherically symmetric background spacetime into even- and odd-mode
perturbations.
In the Part I paper~\cite{K.Nakamura-2021c}, we also gave a strategy
to solve the odd-mode perturbations including $l=0,1$ modes.
Furthermore, we also derived the explicit solutions for the $l=0,1$
odd-mode perturbations to the linearized Einstein equations following
Proposal~\ref{proposal:treatment-proposal-on-pert-on-spherical-BG}.


This paper is the Part II paper of the series of papers on the
application of our gauge-invariant perturbation theory to that on the
Schwarzschild background spacetime.
This series of papers is the full paper version of our short
paper~\cite{K.Nakamura-2021a}.
In this Part II paper, we discuss a strategy to solve the linearized
Einstein equation for even-mode perturbations including $l=0,1$ mode
perturbations.
We also derive the explicit solutions to the $l=0,1$ mode
perturbations with generic linear-order energy-momentum tensor.
As the result, we show that the additional Schwarzschild mass
parameter perturbation in the vacuum case.
This is the realization of the Birkhoff theorem at the
linear-perturbation level in a gauge-invariant manner.
This result is physically reasonable, and it also implies that
Proposal~\ref{proposal:treatment-proposal-on-pert-on-spherical-BG}
is also physically reasonable.
The other supports for
Proposal~\ref{proposal:treatment-proposal-on-pert-on-spherical-BG} are
also given by the realization of exact solutions with matter fields
which will be discussed in the Part III
paper~\cite{K.Nakamura-2021e}.
Furthermore, brief discussions on the extension to the higher-order
perturbations are given in the short paper~\cite{K.Nakamura-2021b}.


The organization of this Part II paper is as follows.
In Sec.~\ref{sec:review-of-general-framework-GI-perturbation-theroy},
after briefly review the framework of the gauge-invariant perturbation
theory, we summarize our proposal in
Refs.~\cite{K.Nakamura-2021a,K.Nakamura-2021c}.
Then, we also summarize the linearized even-mode Einstein equation on
the Schwarzschild background spacetime which was derived in
Ref.~\cite{K.Nakamura-2021c} following
Proposal~\ref{proposal:treatment-proposal-on-pert-on-spherical-BG}.
In Sec.~\ref{sec:Schwarzschild_Background-non-vaccum-even-treatment},
following
Proposal~\ref{proposal:treatment-proposal-on-pert-on-spherical-BG}, we
discuss a strategy to solve these even-mode Einstein equations
including $l=0,1$ mode perturbations.
In Sec.~\ref{sec:l=0_Schwarzschild_Background-non-vac}, we derive the
explicit solutions to the linearized Einstein equation for the
$l=0$ mode perturbations in both the vacuum and the non-vacuum cases.
In Sec.~\ref{sec:l=1_Schwarzschild_Background-non-vac}, we also derive
the explicit solutions to the linearized Einstein equation for the
$l=1$ mode perturbations in both the vacuum and the non-vacuum cases.
The final section (Sec.~\ref{sec:Summary_and_Discussion}) is devoted
to our summary and discussions.


We use the notation used in the previous
papers~\cite{K.Nakamura-2021a,K.Nakamura-2021b,K.Nakamura-2021c}
and the unit $G=c=1$, where $G$ is Newton's constant of gravitation
and $c$ is the velocity of light.


\section{Brief review of the general-relativistic gauge-invariant
  perturbation theory}
\label{sec:review-of-general-framework-GI-perturbation-theroy}


In this section, we review the premise of the series of our
papers~\cite{K.Nakamura-2021a,K.Nakamura-2021c,K.Nakamura-2021e} and
this paper.
In Sec.~\ref{sec:general-framework-GI-perturbation-theroy}, we briefly
review the framework of the gauge-invariant perturbation
theory~\cite{K.Nakamura-2003,K.Nakamura-2005}.
This is an important premise of the series of our papers.
In Sec.~\ref{sec:spherical_background_case}, we review the linear
perturbation on spherically symmetric background spacetimes which
includes our proposal in
Ref.~\cite{K.Nakamura-2021a,K.Nakamura-2021c}.
In Sec.~\ref{sec:Einstein_equations}, we review the linearized
Einstein equations for even-mode perturbations on the Schwarzschild
background spacetime which are to be solved in this paper.


\subsection{General framework of gauge-invariant perturbation theory}
\label{sec:general-framework-GI-perturbation-theroy}


In any perturbation theory, we always treat two spacetime manifolds.
One is the physical spacetime $(\ScrM_{{\rm ph}},\bar{g}_{ab})$,
which is identified with our nature itself, and we want to describe
this spacetime $(\ScrM_{{\rm ph}},\bar{g}_{ab})$ by perturbations.
The other is the background spacetime $(\ScrM,g_{ab})$,
which is prepared as a reference by hand.
Note that these two spacetimes are distinct.
Furthermore, in any perturbation theory, we always write equations
for the perturbation of the variable $Q$ as follows:
\begin{equation}
  \label{eq:variable-symbolic-perturbation}
  Q(``p\mbox{''}) = Q_{0}(p) + \delta Q(p).
\end{equation}
Equation (\ref{eq:variable-symbolic-perturbation}) gives a
relation between variables on different manifolds.
Actually, $Q(``p\mbox{''})$ in
Eq.~(\ref{eq:variable-symbolic-perturbation}) is a variable on
$\ScrM_{\epsilon}=\ScrM_{\rm ph}$, whereas $Q_{0}(p)$ and
$\delta Q(p)$ are variables on $\ScrM$.
Because we regard Eq.~(\ref{eq:variable-symbolic-perturbation}) as
a field equation, Eq.~(\ref{eq:variable-symbolic-perturbation})
includes an implicit assumption of the existence of a point
identification map $\ScrX_{\epsilon}$ $:$
$\ScrM\rightarrow\ScrM_{\epsilon}$ $:$
$p\in\ScrM\mapsto ``p\mbox{''}\in\ScrM_{\epsilon}$.
This identification map is a {\it gauge choice} in
general-relativistic perturbation theories.
This is the notion of the {\it second-kind gauge} pointed out by
Sachs~\cite{R.K.Sachs-1964}.
Note that this second-kind gauge is a different notion from the
degree of freedom of the coordinate transformation on a single
manifold, which is called the {\it first-kind
  gauge}~\cite{K.Nakamura-2010,K.Nakamura-2020,K.Nakamura-2021c}.


To compare with the variable $Q$ on $\ScrM_{\epsilon}$
and its background value $Q_{0}$ on $\ScrM$, we use the pull-back
$\ScrX_{\epsilon}^{*}$ of the identification map
$\ScrX_{\epsilon}$ $:$ $\ScrM\rightarrow\ScrM_{\epsilon}$ and
we evaluate the pulled-back variable $\ScrM_{\epsilon}^{*}Q$ on the
background spacetime $\ScrM$.
Furthermore, in perturbation theories, we expand the pull-back
operation $\ScrX_{\epsilon}^{*}$ to the variable $Q$ with respect
to the infinitesimal parameter $\epsilon$ for the perturbation as
\begin{eqnarray}
  \ScrX_{\epsilon}^{*}Q
  =
  Q_{0}
  + \epsilon {}^{(1)}_{\;\ScrX}Q
  + O(\epsilon^{2})
  .
  \label{eq:perturbative-expansion-of-Q-def}
\end{eqnarray}
Eq.~(\ref{eq:perturbative-expansion-of-Q-def}) are evaluated on the
background spacetime $\ScrM$.
When we have two different gauge choices $\ScrX_{\epsilon}$ and
$\ScrY_{\epsilon}$, we can consider the {\it gauge-transformation},
which is the change of the point-identification
$\ScrX_{\epsilon}\rightarrow\ScrY_{\epsilon}$.
This gauge-transformation is given by the diffeomorphism
$\Phi_{\epsilon}$ $:=$
$\left(\ScrX_{\epsilon}\right)^{-1}\circ\ScrY_{\epsilon}$
$:$ $\ScrM$ $\rightarrow$ $\ScrM$.
Actually, the diffeomorphism $\Phi_{\epsilon}$ induces a pull-back from
the representation $\ScrX_{\epsilon}^{*}\!Q_{\epsilon}$ to the
representation $\ScrY_{\epsilon}^{*}\!Q_{\epsilon}$ as
$\ScrY_{\epsilon}^{*}\!Q_{\epsilon}=\Phi_{\epsilon}^{*}\ScrX_{\epsilon}^{*}\!Q_{\epsilon}$.
From general arguments of the Taylor
expansion\cite{M.Bruni-S.Matarrese-S.Mollerach-S.Sonego-1997}, the
pull-back $\Phi_{\epsilon}^{*}$ is expanded as
\begin{eqnarray}
  \ScrY_{\epsilon}^{*}\!Q_{\epsilon}
  &=&
  \ScrX_{\epsilon}^{*}\!Q_{\epsilon}
  + \epsilon {\pounds}_{\xi_{(1)}} \ScrX_{\epsilon}^{*}\!Q_{\epsilon}
  + O(\epsilon^{2}),
  \label{eq:Bruni-46-one}
\end{eqnarray}
where $\xi_{(1)}^{a}$ is the generator of $\Phi_{\epsilon}$.
From Eqs.~(\ref{eq:perturbative-expansion-of-Q-def}) and
(\ref{eq:Bruni-46-one}), the gauge-transformation for the first-order
perturbation ${}^{(1)}Q$ is given by
\begin{eqnarray}
  \label{eq:Bruni-47-one}
  {}^{(1)}_{\;\ScrY}\!Q - {}^{(1)}_{\;\ScrY}\!Q &=&
  {\pounds}_{\xi_{(1)}}Q_{0}.
\end{eqnarray}
We also employ the {\it order by order gauge invariance} as a
concept of gauge invariance~\cite{K.Nakamura-2009a}.
We call the $k$th-order perturbation ${}^{(k)}_{\ScrX}\!Q$ as
gauge invariant if and only if
\begin{eqnarray}
  \label{eq:notion-def-gauge-inv}
  {}^{(k)}_{\;\ScrX}\!Q = {}^{(k)}_{\;\ScrY}\!Q
\end{eqnarray}
for any gauge choice $\ScrX_{\epsilon}$ and $\ScrY_{\epsilon}$.


Based on the above setup, we proposed a procedure to construct
gauge-invariant variables of higher-order
perturbations~\cite{K.Nakamura-2003,K.Nakamura-2005}.
First, we expand the metric on the physical spacetime
$\ScrM_{\epsilon}$, which was pulled back to the background
spacetime $\ScrM$ through a gauge choice $\ScrX_{\epsilon}$ as
\begin{eqnarray}
  \ScrX^{*}_{\epsilon}\bar{g}_{ab}
  &=&
  g_{ab} + \epsilon {}_{\ScrX}\!h_{ab}
  + O(\epsilon^{2}).
  \label{eq:metric-expansion}
\end{eqnarray}
Although the expression (\ref{eq:metric-expansion}) depends
entirely on the gauge choice $\ScrX_{\epsilon}$, henceforth,
we do not explicitly express the index of the gauge choice
$\ScrX_{\epsilon}$ in the expression if there is no
possibility of confusion.
The important premise of our proposal was the following
conjecture~\cite{K.Nakamura-2003,K.Nakamura-2005} for the linear metric
perturbation $h_{ab}$:
\begin{conjecture}
  \label{conjecture:decomposition-conjecture}
  If the gauge-transformation rule for a perturbative pulled-back
  tensor field $h_{ab}$ to the background spacetime $\ScrM$ is
  given by ${}_{\ScrY}\!h_{ab}$ $-$ ${}_{\ScrX}\!h_{ab}$ $=$
  ${\pounds}_{\xi_{(1)}}g_{ab}$ with the background metric $g_{ab}$,
  there then exist a tensor field $\ScrF_{ab}$ and a vector field
  $Y^{a}$ such that $h_{ab}$ is decomposed as $h_{ab}$ $=:$
  $\ScrF_{ab}$ $+$ ${\pounds}_{Y}g_{ab}$, where $\ScrF_{ab}$ and
  $Y^{a}$ are transformed as ${}_{\ScrY}\!\ScrF_{ab}$ $-$
  ${}_{\ScrX}\!\ScrF_{ab}$ $=$ $0$ and ${}_{\ScrY}\!Y^{a}$
  $-$ ${}_{\ScrX}\!Y^{a}$ $=$ $\xi^{a}_{(1)}$ under the gauge
  transformation, respectively.
\end{conjecture}
We call $\ScrF_{ab}$ and $Y^{a}$ as the
{\it gauge-invariant} and {\it gauge-variant} parts
of $h_{ab}$, respectively.


The proof of Conjecture~\ref{conjecture:decomposition-conjecture} is
highly nontrivial~\cite{K.Nakamura-2011,K.Nakamura-2013}, and it was
found that the gauge-invariant variables are essentially non-local.
Despite this non-triviality, once we accept
Conjecture~\ref{conjecture:decomposition-conjecture},
we can decompose the linear perturbation of an arbitrary tensor field
${}_{\ScrX}^{(1)}\!Q$, whose gauge-transformation is given by
Eq.~(\ref{eq:Bruni-47-one}), through the gauge-variant part $Y_{a}$ of
the metric perturbation in
Conjecture~\ref{conjecture:decomposition-conjecture} as
\begin{eqnarray}
  \label{eq:arbitrary-Q-decomp}
  {}_{\ScrX}^{(1)}\!Q = {}^{(1)}\!\ScrQ + {\pounds}_{{}_{\ScrX}\!Y}Q_{0}.
\end{eqnarray}


As examples, the linearized Einstein tensor
${}_{\ScrX}^{(1)}G_{a}^{\;\;b}$ and the linear perturbation of the
energy-momentum tensor ${}_{\ScrX}^{(1)}T_{a}^{\;\;b}$ are also
decomposed as
\begin{eqnarray}
  \label{eq:Gab-Tab-decomp}
  {}_{\ScrX}^{(1)}\!G_{a}^{\;\;b}
  =
  {}^{(1)}\!\ScrG_{a}^{\;\;b}\left[\ScrF\right] + {\pounds}_{{}_{\ScrX}\!Y}G_{a}^{\;\;b}
  ,
  \quad
  {}_{\ScrX}^{(1)}\!T_{a}^{\;\;b}
  =
  {}^{(1)}\!\ScrT_{a}^{\;\;b} + {\pounds}_{{}_{\ScrX}\!Y}T_{a}^{\;\;b}
  ,
\end{eqnarray}
where $G_{ab}$ and $T_{ab}$ are the background values of the Einstein
tensor and the energy-momentum tensor, respectively.
The gauge-invariant part ${}^{(1)}\!\ScrG_{a}^{\;\;b}$ of the
linear-order perturbation of the Einstein tensor is given by
\begin{eqnarray}
  \label{eq:linear-Einstein-AIA2010-2}
  \!\!\!\!\!\!\!\!\!\!\!\!\!\!\!\!
  &&
     {}^{(1)}\ScrG_{a}^{\;\;b}\left[A\right]
     :=
     {}^{(1)}\Sigma_{a}^{\;\;b}\left[A\right]
     - \frac{1}{2} \delta_{a}^{\;\;b} {}^{(1)}\Sigma_{c}^{\;\;c}\left[A\right]
     ,
  \\
  \label{eq:(1)Sigma-def-linear}
  \!\!\!\!\!\!\!\!\!\!\!\!\!\!\!\!
  &&
     {}^{(1)}\Sigma_{a}^{\;\;b}\left[A\right]
     :=
     - 2 \nabla_{[a}^{}H_{d]}^{\;\;\;bd}\left[A\right]
     - A^{cb} R_{ac}
     , \quad
     H_{ba}^{\;\;\;\;c}\left[A\right]
     :=
     \nabla_{(a}A_{b)}^{\;\;c} - \frac{1}{2} \nabla^{c}A_{ab}
     ,
\end{eqnarray}
where $A_{ab}$ is an arbitrary tensor field of the second rank.
Then, using the background Einstein equation
$G_{a}^{\;\;b}=8\pi T_{a}^{\;\;b}$, the linearized Einstein equation
${}_{\ScrX}^{(1)}\!G_{ab}=8\pi{}_{\ScrX}^{(1)}\!T_{ab}$ is
automatically given in the gauge-invariant form
\begin{eqnarray}
  \label{eq:einstein-equation-gauge-inv}
  {}^{(1)}\!\ScrG_{a}^{\;\;b}\left[\ScrF\right] = 8 \pi {}^{(1)}\!\ScrT_{a}^{\;\;b}
\end{eqnarray}
even if the background Einstein equation is nontrivial.
We also note that, in the case of a vacuum background case, i.e.,
$G_{a}^{\;\;b} =8\pi T_{a}^{\;\;b}=0$, Eq.~(\ref{eq:Gab-Tab-decomp}) shows that the
linear perturbations of the Einstein tensor and the energy-momentum
tensor is automatically gauge-invariant of the second kind.


We can also derive the perturbation of the divergence of
$\bar{\nabla}_{a}\bar{T}_{b}^{\;\;a}$ of the second-rank tensor
$\bar{T}_{b}^{\;\;a}$ on $(\ScrM_{\rm ph},\bar{g}_{ab})$.
Through the gauge choice $\ScrX_{\epsilon}$, $\bar{T}_{b}^{\;\;a}$
is pulled back to $\ScrX_{\epsilon}^{*}\bar{T}_{b}^{\;\;a}$ on the
background spacetime $(\ScrM,g_{ab})$, and the
covariant derivative operator $\bar{\nabla}_{a}$ on
$(\ScrM_{\rm ph},\bar{g}_{ab})$ is pulled back to a derivative
operator
$\bar{\nabla}_{a}(=\ScrX_{\epsilon}^{*}\bar{\nabla}_{a}(\ScrX_{\epsilon}^{-1})^{*})$
on $(\ScrM,g_{ab})$.
Note that the derivative $\bar{\nabla}_{a}$ is the covariant
derivative associated with the metric
$\ScrX_{\epsilon}\bar{g}_{ab}$, whereas the derivative $\nabla_{a}$ on
the background spacetime $(\ScrM,g_{ab})$ is the covariant derivative
associated with the background metric $g_{ab}$.
Bearing in mind the difference in these derivatives, the first-order
perturbation of $\bar{\nabla}_{a}\bar{T}_{b}^{\;\;a}$ is given by
\begin{eqnarray}
  \label{eq:linear-perturbation-of-div-Tab}
  {}^{(1)}\!\left(\bar{\nabla}_{a}\bar{T}_{b}^{\;\;a}\right)
  =
  \nabla_{a}{}^{(1)}\!\ScrT_{b}^{\;\;a}
  +
  H_{ca}^{\;\;\;\;a}\left[\ScrF\right] T_{b}^{\;\;c}
  -
  H_{ba}^{\;\;\;\;c}\left[\ScrF\right] T_{c}^{\;\;a}
  +
  {\pounds}_{Y}\nabla_{a}T_{b}^{\;\;a}
  .
\end{eqnarray}
The derivation of the formula
(\ref{eq:linear-perturbation-of-div-Tab}) is given in
Ref.~\cite{K.Nakamura-2005}.
If the tensor field $\bar{T}_{b}^{\;\;a}$ is the Einstein tensor
$\bar{G}_{a}^{\;\;b}$, Eq.~(\ref{eq:linear-perturbation-of-div-Tab})
yields the linear-order perturbation of the Bianchi identity
\begin{eqnarray}
  \label{eq:linear-perturbation-of-div-Gab}
  \nabla_{a}{}^{(1)}\!\ScrG_{b}^{\;\;a}\left[\ScrF\right]
  +
  H_{ca}^{\;\;\;\;a}\left[\ScrF\right] G_{b}^{\;\;c}
  -
  H_{ba}^{\;\;\;\;c}\left[\ScrF\right] G_{c}^{\;\;a}
  =
  0
\end{eqnarray}
and if the background Einstein tensor vanishes $G_{a}^{\;\;b}=0$, we
obtain the identity
\begin{eqnarray}
  \label{eq:linear-perturbation-of-div-Gab-vacuum}
  \nabla_{a}{}^{(1)}\!\ScrG_{b}^{\;\;a}\left[\ScrF\right]
  =
  0.
\end{eqnarray}
By contrast, if the tensor field $\bar{T}_{b}^{\;\;a}$ is the
energy-momentum tensor, Eq.~(\ref{eq:linear-perturbation-of-div-Tab})
yields the continuity equation of the energy-momentum tensor
\begin{eqnarray}
  \label{eq:linear-perturbation-of-div-Tab-ene-mon}
  \nabla_{a}{}^{(1)}\!\ScrT_{b}^{\;\;a}
  +
  H_{ca}^{\;\;\;\;a}\left[\ScrF\right] T_{b}^{\;\;c}
  -
  H_{ba}^{\;\;\;\;c}\left[\ScrF\right] T_{c}^{\;\;a}
  =
  0
  ,
\end{eqnarray}
where we used the background continuity equation
$\nabla_{a}T_{b}^{\;\;a}=0$.
If the background spacetime is vacuum $T_{ab}=0$,
Eq.~(\ref{eq:linear-perturbation-of-div-Tab-ene-mon}) yields
a linear perturbation of the energy-momentum tensor given by
\begin{eqnarray}
  \label{eq:divergence-barTab-linear-vac-back-u}
  \nabla_{a}{}^{(1)}\!\ScrT_{b}^{\;\;a}
  =
  0
  .
\end{eqnarray}


We should note that the decomposition of the metric perturbation
$h_{ab}$ into its gauge-invariant part $\ScrF_{ab}$ and
into its gauge-variant part $Y^{a}$ is not
unique~\cite{K.Nakamura-2009a,K.Nakamura-2010,K.Nakamura-2020}.
As explained in the Part I paper~\cite{K.Nakamura-2021c}, for example,
the gauge-invariant part $\ScrF_{ab}$ has six components and we can
create the gauge-invariant vector field $Z^{a}$ through these
components of the gauge-invariant metric perturbation $\ScrF_{ab}$
such that the gauge-transformation of the vector field $Z^{a}$ is
given by ${}_{\ScrY}\!Z^{a}$ $-$ ${}_{\ScrX}\!Z^{a}$ $=$ $0$.
Using this gauge-invariant vector field $Z^{a}$, the original metric
perturbation can be expressed as follows:
\begin{eqnarray}
  \label{eq:gauge-inv-nonunique}
  h_{ab}
  =
  \ScrF_{ab} - {\pounds}_{Z}g_{ab}
  + {\pounds}_{Z+Y}g_{ab}
  =:
  \ScrH_{ab} + {\pounds}_{X}g_{ab}
  .
\end{eqnarray}
The tensor field $\ScrH_{ab}:=\ScrF_{ab} - {\pounds}_{Z}g_{ab}$
is also regarded as the gauge-invariant part of the perturbation
$h_{ab}$ because
${}_{\ScrY}\!\ScrH_{ab}-{}_{\ScrX}\!\ScrH_{ab}=0$.
Similarly, the vector field $X^{a}:=Z^{a}+Y^{a}$ is also regarded as
the gauge-variant part of the perturbation $h_{ab}$ because
${}_{\ScrY}\!X^{a}$ $-$ ${}_{\ScrX}\!X^{a}$ $=$ $\xi^{a}_{(1)}$.
This non-uniqueness appears in the solutions derived in
Secs.~\ref{sec:l=0_Schwarzschild_Background-non-vac}
and~\ref{sec:l=1_Schwarzschild_Background-non-vac}, as in the case of
the $l=1$ odd-mode perturbative solutions in the Part I
paper~\cite{K.Nakamura-2021c}.
These non-uniqueness of gauge-invariant variable can be regarded as
the first-kind gauge as explained in Part I
paper~\cite{K.Nakamura-2021c}, i.e., degree of freedom of the choice
of the coordinate system on the physical spacetime
$\ScrM_{\epsilon}$.
Since we often use the first-kind gauge when we predict and interpret
the measurement results of observations and experiments, we should
regard that this non-uniqueness of gauge-invariant variable of the
second kind may have some physical meaning~\cite{K.Nakamura-2021c}.


\subsection{Linear perturbations on spherically symmetric background}
\label{sec:spherical_background_case}


Here, we consider the 2+2 formulation of the perturbation of a
spherically symmetric background spacetime, which originally proposed
by Gerlach and Sengupta~\cite{Gerlach_Sengupta-1979a,Gerlach_Sengupta-1979b,Gerlach_Sengupta-1979c,Gerlach_Sengupta-1980}.
Spherically symmetric spacetimes are characterized by the direct
product $\ScrM=\ScrM_{1}\times S^{2}$ and their metric is
\begin{eqnarray}
  \label{eq:background-metric-2+2}
  g_{ab}
  &=&
  y_{ab} + r^{2}\gamma_{ab}
  , \\
  y_{ab} &=& y_{AB} (dx^{A})_{a}(dx^{B})_{b}, \quad
             \gamma_{ab} = \gamma_{pq} (dx^{p})_{a} (dx^{q})_{b},
\end{eqnarray}
where $x^{A} = (t,r)$, $x^{p}=(\theta,\phi)$, and $\gamma_{pq}$ is the
metric on the unit sphere.
In the Schwarzschild spacetime, the metric
(\ref{eq:background-metric-2+2}) is given by
\begin{eqnarray}
  \label{eq:background-metric-2+2-y-comp-Schwarzschild}
  y_{ab}
  &=&
      - f (dt)_{a}(dt)_{b}
      +
      f^{-1} (dr)_{a}(dr)_{b}
      ,
      \quad
      f = 1 - \frac{2M}{r}
  ,\\
  \label{eq:background-metric-2+2-gamma-comp-Schwarzschild}
  \gamma_{ab}
  &=&
      (d\theta)_{a}(d\theta)_{b}
      +
      \sin^{2}\theta(d\phi)_{a}(d\phi)_{b}
      =
      \theta_{a}\theta_{b} + \phi_{a}\phi_{b}
      ,
  \\
  \label{eq:S2-unit-basis-def}
  \theta_{a}
  &=&
      (d\theta)_{a}, \quad
      \phi_{a}
      =
      \sin\theta (d\phi)_{a}
      .
\end{eqnarray}


On this background spacetime $(\ScrM,g_{ab})$, the components of
the metric perturbation are given by
\begin{eqnarray}
  \label{eq:metric-perturbation-components}
  h_{ab}
  =
  h_{AB} (dx^{A})_{a}(dx^{B})_{b}
  +
  2 h_{Ap} (dx^{A})_{(a}(dx^{p})_{b)}
  +
  h_{pq} (dx^{p})_{a}(dx^{q})_{b}
  .
\end{eqnarray}
Here, we note that the components $h_{AB}$, $h_{Ap}$, and
$h_{pq}$ are regarded as components of scalar, vector, and
tensor on $S^{2}$, respectively.
In many literatures, these components are decomposed through the
decomposition~\cite{J.W.York-1973,J.W.York-1974,S.Deser-1967} using
the spherical harmonics $S=Y_{lm}$ as follows:
\begin{eqnarray}
  \label{eq:hAB-fourier}
  h_{AB}
  \!\!\!\!&=&\!\!\!\!
      \sum_{l,m} \tilde{h}_{AB} S
      ,
  \\
  \label{eq:hAp-fourier}
  h_{Ap}
  \!\!\!\!&=&\!\!\!\!
      r \sum_{l,m} \left[
      \tilde{h}_{(e1)A} \hat{D}_{p}S
      +
      \tilde{h}_{(o1)A} \epsilon_{pq} \hat{D}^{q}S
      \right]
      ,
  \\
  \label{eq:hpq-fourier}
  h_{pq}
  \!\!\!\!&=&\!\!\!\!
      r^{2} \sum_{l,m} \left[
      \frac{1}{2} \gamma_{pq} \tilde{h}_{(e0)} S
      +
      \tilde{h}_{(e2)} \left(
      \hat{D}_{p}\hat{D}_{q} - \frac{1}{2} \gamma_{pq} \hat{D}^{r}\hat{D}_{r}
      \right) S
      +
      2 \tilde{h}_{(o2)} \epsilon_{r(p} \hat{D}_{q)}\hat{D}^{r} S
      \right]
      ,
\end{eqnarray}
where $\hat{D}_{p}$ is the covariant derivative associated with
the metric $\gamma_{pq}$ on $S^{2}$,
$\hat{D}^{p}=\gamma^{pq}\hat{D}_{q}$,
$\epsilon_{pq}=\epsilon_{[pq]}=2\theta_{[p}\phi_{q]}$ is the totally
antisymmetric tensor on $S^{2}$.


If we employ the decomposition
(\ref{eq:hAB-fourier})--(\ref{eq:hpq-fourier}) with $S=Y_{lm}$ to the
metric perturbation $h_{ab}$, special treatments for $l=0,1$ modes are
required~\cite{T.Regge-J.A.Wheeler-1957,F.Zerilli-1970-PRL,F.Zerilli-1970-PRD,H.Nakano-2019,V.Moncrief-1974a,V.Moncrief-1974b,C.T.Cunningham-R.H.Price-V.Moncrief-1978,Chandrasekhar-1983,Gerlach_Sengupta-1979a,Gerlach_Sengupta-1979b,Gerlach_Sengupta-1979c,Gerlach_Sengupta-1980,T.Nakamura-K.Oohara-Y.Kojima-1987,Gundlach-Martine-Garcia-2000,Gundlach-Martine-Garcia-2001,A.Nagar-L.Rezzolla-2005-2006,K.Martel-E.Poisson-2005}.
This is due to the fact that the set of harmonic functions
\begin{eqnarray}
  \label{eq:harmonic-fucntions-set}
  \left\{
  S,
  \hat{D}_{p}S,
  \epsilon_{pq}\hat{D}^{q}S,
  \frac{1}{2}\gamma_{pq}S,
  \left(\hat{D}_{p}\hat{D}_{q}-\frac{1}{2}\gamma_{pq}\right)S,
  2\epsilon_{r(p}\hat{D}_{q)}\hat{D}^{r}S
  \right\}
\end{eqnarray}
loses its linear-independence for $l=0,1$ modes.
Actually, the inverse-relation of the decomposition formulae
(\ref{eq:hAB-fourier})--(\ref{eq:hpq-fourier}) requires the Green
functions of the derivative operators
$\hat{\Delta}:=\hat{D}^{r}\hat{D}_{r}$ and
$\hat{\Delta}+2:=\hat{D}^{r}\hat{D}_{r}+2$.
Since the eigen modes of these operators are $l=0$ and $l=1$,
respectively, this is the reason why the special treatments for these
modes are required.
However, these special treatments become an obstacle when we develop
higher-order perturbation
theory~\cite{D.Brizuela-J.M.Martin-Garcia-G.A.M.Marugan-2007}.


To resolve this $l=0,1$ mode problem, in Part I
paper~\cite{K.Nakamura-2021a,K.Nakamura-2021c}, we chose the scalar
function $S$ as
\begin{eqnarray}
  \label{eq:harmonics-extended-choice-sum}
  S
  =
  S_{\delta}
  =
  \left\{
  \begin{array}{ccccc}
    Y_{lm} & \quad & \mbox{for} & \quad & l\geq 2; \\
    k_{(\hat{\Delta}+2)m} & \quad & \mbox{for} & \quad & l=1; \\
    k_{(\hat{\Delta})} & \quad & \mbox{for} & \quad & l=0.
  \end{array}
  \right.
\end{eqnarray}
and use the decomposition formulae
(\ref{eq:hAB-fourier})--(\ref{eq:hpq-fourier}), where the functions
$k_{(\hat{\Delta})}$ and $k_{(\hat{\Delta}+2)}$ satisfy the equation
\begin{eqnarray}
  \label{eq:kernel-mode-def-l=0-l=1}
  \hat{\Delta} k_{(\hat{\Delta})} = 0, \quad
  \left(\hat{\Delta} + 2 \right) k_{(\hat{\Delta}+2)} = 0,
\end{eqnarray}
respectively.
As shown in Part I paper~\cite{K.Nakamura-2021c}, the set of harmonic
functions (\ref{eq:harmonic-fucntions-set}) becomes the
linear-independent set including $l=0,1$ modes if we employ
\begin{eqnarray}
  k_{(\hat{\Delta})}
  &=&
      1 + \delta \ln\left(\frac{1-\cos\theta}{1+\cos\theta}\right)^{1/2},
      \quad \delta \in\RF
      ,
      \label{eq:l=0-general-mode-func-specific}
  \\
  k_{(\hat{\Delta}+2,m=0)}
  &=&
      \cos\theta
      +
      \delta \left(\frac{1}{2} \cos\theta \ln\frac{1+\cos\theta}{1-\cos\theta} -1\right)
      ,
     \quad \delta \in \RF
     ,
     \label{eq:l=1-m=0-mode-func-explicit}
     \\
  k_{(\hat{\Delta}+2,m=\pm1)}
  &=&
      \left[
      \sin\theta
      +
      \delta \left(
      + \frac{1}{2} \sin\theta \ln\frac{1+\cos\theta}{1-\cos\theta}
      + \cot\theta
      \right)
      \right]
      e^{\pm i\phi}
      .
      \label{eq:l=1-m=pm1-mode-func-explicit}
\end{eqnarray}
These choices guarantee the one-to-one correspondence between the
components $\{h_{AB},$ $h_{Ap},$ $h_{pq}\}$ and the mode coefficients
$\{\tilde{h}_{AB},$ $\tilde{h}_{(e1)A},$ $\tilde{h}_{(o1)A},$
$\tilde{h}_{(e0)},$ $\tilde{h}_{(e2)},$ $\tilde{h}_{(o2)}\}$ with the
decomposition formulae (\ref{eq:hAB-fourier})-(\ref{eq:hpq-fourier})
owing to the linear-independence of the set of the harmonic functions
(\ref{eq:harmonic-fucntions-set}) when $\delta\neq 0$.
Then, the mode-by-mode analysis including $l=0,1$ is possible when
$\delta\neq 0$.
However, the mode functions
(\ref{eq:l=0-general-mode-func-specific})--(\ref{eq:l=1-m=pm1-mode-func-explicit})
are singular if $\delta\neq 0$.
When $\delta=0$, we have $k_{(\hat{\Delta})}\propto Y_{00}$ and
$k_{(\hat{\Delta}+2)m}\propto Y_{1m}$.
Using the above harmonics functions $S_{\delta}$ in
Eq.~(\ref{eq:harmonics-extended-choice-sum}), we propose the
following strategy:
\begin{proposal}
  \label{proposal:treatment-proposal-on-pert-on-spherical-BG}
  We decompose the metric perturbation $h_{ab}$ on the background
  spacetime with the metric
  (\ref{eq:background-metric-2+2})--(\ref{eq:background-metric-2+2-gamma-comp-Schwarzschild})
  through Eqs.~(\ref{eq:hAB-fourier})--(\ref{eq:hpq-fourier}) with the
  harmonic function $S_{\delta}$ given by
  Eq.~(\ref{eq:harmonics-extended-choice-sum}).
  Then, Eqs.~(\ref{eq:hAB-fourier})--(\ref{eq:hpq-fourier}) become
  invertible including $l=0,1$ modes.
  After deriving the mode-by-mode field equations such as linearized
  Einstein equations by using the harmonic functions $S_{\delta}$, we
  choose $\delta=0$ as regular boundary condition for solutions when
  we solve these field equations.
\end{proposal}


As shown in the Part I paper~\cite{K.Nakamura-2021c}, once we accept
Proposal~\ref{proposal:treatment-proposal-on-pert-on-spherical-BG},
the Conjecture~\ref{conjecture:decomposition-conjecture} becomes the
following statement:
\begin{theorem}
  \label{theorem:decomposition-theorem-with-spherical-symmetry}
  If the gauge-transformation rule for a perturbative pulled-back
  tensor field $h_{ab}$ to the background spacetime $\ScrM$ is
  given by ${}_{\ScrY}\!h_{ab}$ $-$ ${}_{\ScrX}\!h_{ab}$ $=$
  ${\pounds}_{\xi_{(1)}}g_{ab}$ with the background metric $g_{ab}$
  with spherical symmetry, there then exist a tensor field
  $\ScrF_{ab}$ and a vector field $Y^{a}$ such that $h_{ab}$ is
  decomposed as $h_{ab}$ $=:$ $\ScrF_{ab}$ $+$
  ${\pounds}_{Y}g_{ab}$, where $\ScrF_{ab}$ and $Y^{a}$ are
  transformed as ${}_{\ScrY}\!\ScrF_{ab}$ $-$
  ${}_{\ScrX}\!\ScrF_{ab}$ $=$ $0$ and ${}_{\ScrY}\!Y^{a}$
  $-$ ${}_{\ScrX}\!Y^{a}$ $=$ $\xi^{a}_{(1)}$ under the gauge
  transformation, respectively.
\end{theorem}
Furthermore, including $l=0,1$ modes, the components of the
gauge-invariant part $\ScrF_{ab}$ of the metric perturbation
$h_{ab}$ is given by
\begin{eqnarray}
  \label{eq:2+2-gauge-invariant-variables-calFAB}
  \ScrF_{AB}
  &=&
      \sum_{l,m} \tilde{F}_{AB} S_{\delta}
      ,
  \\
  \label{eq:2+2-gauge-invariant-variables-calFAp}
  \ScrF_{Ap}
  &=&
      r \sum_{l,m} \tilde{F}_{A} \epsilon_{pq}
      \hat{D}^{q}S_{\delta}, \quad
      \hat{D}^{p}\ScrF_{Ap} = 0
      ,
  \\
  \label{eq:2+2-gauge-invariant-variables-calFpq}
  \ScrF_{pq}
  &=&
      = \frac{1}{2} \gamma_{pq} r^{2} \sum_{l,m} \tilde{F} S_{\delta}
      .
\end{eqnarray}
Thus, we have resolved the zero-mode problem in the perturbations on
the spherically symmetric background spacetimes.
Through the gauge-invariant variables
(\ref{eq:2+2-gauge-invariant-variables-calFAB})--(\ref{eq:2+2-gauge-invariant-variables-calFpq}),
we derived the linearized Einstein equations in Part I
paper~\cite{K.Nakamura-2021c}.


\subsection{Even-mode linearized Einstein equations}
\label{sec:Einstein_equations}


Since the odd-mode perturbations are discussed in Part I
paper~\cite{K.Nakamura-2021c}, we consider the linearized even-mode
Einstein equations on the Schwarzschild background spacetime in this
paper.
The Schwarzschild spacetime is vacuum solution to the Einstein
equation $G_{a}^{\;\;b}=0=T_{a}^{\;\;b}$.
Since we proved
Theorem~\ref{theorem:decomposition-theorem-with-spherical-symmetry} on
the spherically symmetric background spacetime, the linearized
Einstein equation is given in the following gauge-invariant
form as Eq.~(\ref{eq:einstein-equation-gauge-inv}).
To evaluate the Einstein equation
(\ref{eq:einstein-equation-gauge-inv}) through the mode-by-mode
analysis including $l=0,1$, we also consider the mode-decomposition of
the gauge-invariant part ${}^{(1)}\!\ScrT_{ab}$ $:=$
$g_{bc}{}^{(1)}\!\ScrT_{a}^{\;\;c}$ of the linear-perturbation of
the energy momentum tensor through the set
(\ref{eq:harmonic-fucntions-set}) of the harmonics as follows:
\begin{eqnarray}
  {}^{(1)}\!\ScrT_{ab}
  &=&
      \sum_{l,m}
      \tilde{T}_{AB}
      S_{\delta}
      (dx^{A})_{a}(dx^{B})_{b}
      +
      r
      \sum_{l,m} \left\{
      \tilde{T}_{(e1)A} \hat{D}_{p}S_{\delta}
      +
      \tilde{T}_{(o1)A} \epsilon_{pr}\hat{D}^{r}S_{\delta}
      \right\}
      2 (dx^{A})_{(a}(dx^{p})_{b)}
      \nonumber\\
  &&
     +
     \sum_{l,m} \left\{
     \tilde{T}_{(e0)} \frac{1}{2} \gamma_{pq} S_{\delta}
     +
     \tilde{T}_{(e2)} \left(
     \hat{D}_{p}\hat{D}_{q}S_{\delta}
     -
     \frac{1}{2} \gamma_{pq} \hat{D}_{r}\hat{D}^{r}S_{\delta}
     \right)
     \right.
     \nonumber\\
  && \quad\quad\quad
     \left.
     +
     \tilde{T}_{(o2)}
     2 \epsilon_{r(p}\hat{D}_{q)}\hat{D}^{r}S_{\delta}
     \right\}
     (dx^{p})_{a}(dx^{q})_{b}
     .
     \label{eq:1st-pert-calTab-dd-decomp}
\end{eqnarray}


Since the background spacetime is vacuum, the pull-backed divergence
of the energy-momentum tensor is given by
Eq.~(\ref{eq:divergence-barTab-linear-vac-back-u}) and the even-mode
components of Eq.~(\ref{eq:divergence-barTab-linear-vac-back-u}) in
terms of the mode coefficients defined by
Eq.~(\ref{eq:1st-pert-calTab-dd-decomp}) are given by
\begin{eqnarray}
  &&
     \bar{D}^{C}\tilde{T}_{C}^{\;\;B}
     + \frac{2}{r} (\bar{D}^{D}r)\tilde{T}_{D}^{\;\;\;B}
     -  \frac{1}{r} l(l+1) \tilde{T}_{(e1)}^{B}
     -  \frac{1}{r} (\bar{D}^{B}r) \tilde{T}_{(e0)}
     =
     0
     ,
     \label{eq:div-barTab-linear-vac-back-u-A-mode}
  \\
  &&
     \bar{D}^{C}\tilde{T}_{(e1)C}
     + \frac{3}{r} (\bar{D}^{C}r) \tilde{T}_{(e1)C}
     + \frac{1}{2r} \tilde{T}_{(e0)}
     -  \frac{1}{2r} (l-1)(l+2) \tilde{T}_{(e2)}
     =
     0
     .
     \label{eq:div-barTab-linear-vac-back-u-p-mode-even}
\end{eqnarray}


Owing to the linear-independence of the set
(\ref{eq:harmonic-fucntions-set}) of the harmonics, we can evaluate
the gauge-invariant linearized Einstein equation
(\ref{eq:einstein-equation-gauge-inv}) through the mode-by-mode
analyses including $l=0,1$ modes.
As summarized in the Part I paper~\cite{K.Nakamura-2021c}, the
traceless even part of the $(p,q)$-component of the linearized
Einstein equation (\ref{eq:einstein-equation-gauge-inv}) is given by
\begin{eqnarray}
  \label{eq:linearized-Einstein-pq-traceless-even}
  \tilde{F}_{D}^{\;\;\;D}
  =
  -
  16 \pi r^{2}
  \tilde{T}_{(e2)}
  .
\end{eqnarray}
Using this equation, the even part of $(A,q)$-component, equivalently
$(p,B)$-component, of the linearized Einstein equation
(\ref{eq:einstein-equation-gauge-inv}) yields
\begin{eqnarray}
  \bar{D}^{D}\tilde{\FF}_{AD}
  - \frac{1}{2} \bar{D}_{A}\tilde{F}
  =
  16 \pi
  \left[
  r \tilde{T}_{(e1)A}
  - \frac{1}{2} r^{2} \bar{D}_{A}\tilde{T}_{(e2)}
  \right]
  =:
  16 \pi S_{(ec)A}
  \label{eq:even-FAB-divergence}
\end{eqnarray}
through the definition of the traceless part $\tilde{\FF}_{AB}$ of
the variable $\tilde{F}_{AB}$:
\begin{eqnarray}
  \label{eq:FF-def}
  \tilde{\FF}_{AB} := \tilde{F}_{AB} - \frac{1}{2} y_{AB} \tilde{F}_{C}^{\;\;C}.
\end{eqnarray}


Using Eqs.~(\ref{eq:linearized-Einstein-pq-traceless-even}),
(\ref{eq:even-FAB-divergence}), and the background Einstein,
the trace part of $(p,q)$-component of the linearized Einstein
equation (\ref{eq:einstein-equation-gauge-inv}) yields
Eq.~(\ref{eq:div-barTab-linear-vac-back-u-p-mode-even}).


Finally, through Eqs.~(\ref{eq:linearized-Einstein-pq-traceless-even})
and (\ref{eq:even-FAB-divergence}) and the background Einstein
equations, the trace part of the $(A,B)$-component of the linearized
Einstein equation (\ref{eq:einstein-equation-gauge-inv}) is given by
\begin{eqnarray}
  &&
     \!\!\!\!\!\!\!\!\!\!\!\!\!\!\!\!\!\!\!\!\!\!\!\!
     \left(
     \bar{D}_{D}\bar{D}^{D}
     + \frac{2}{r} (\bar{D}^{D}r) \bar{D}_{D}
     -  \frac{(l-1)(l+2)}{r^{2}}
     \right) \tilde{F}
     - \frac{4}{r^{2}} (\bar{D}_{C}r) (\bar{D}_{D}r) \tilde{\FF}^{CD}
  =
      16 \pi S_{(F)}
      ,
     \label{eq:even-mode-tildeF-master-eq}
  \\
  S_{(F)}
  \!\!\!\!&:=&\!\!\!\!
       \tilde{T}_{C}^{\;\;\;C}
       + 4 (\bar{D}_{D}r) \tilde{T}_{(e1)}^{D}
       -  2 r (\bar{D}_{D}r) \bar{D}^{D}\tilde{T}_{(e2)}
       -  (l(l+1)+2) \tilde{T}_{(e2)}
       .
               \label{eq:sourcee0-def}
\end{eqnarray}
On the other hand, the traceless part of the $(A,B)$-component of the
linearized Einstein equation (\ref{eq:einstein-equation-gauge-inv}) is
given by
\begin{eqnarray}
  &&
     \!\!\!\!\!\!\!\!\!\!\!\!\!\!\!\!\!\!\!\!\!\!\!\!
     \left[
     -  \bar{D}_{D}\bar{D}^{D}
     -  \frac{2}{r} (\bar{D}_{D}r) \bar{D}^{D}
     + \frac{4}{r} (\bar{D}^{D}\bar{D}_{D}r)
     + \frac{l(l+1)}{r^{2}}
     \right]
     \tilde{\FF}_{AB}
     \nonumber\\
  \!\!\!\!&&\!\!\!\!
     + \frac{4}{r} (\bar{D}^{D}r) \bar{D}_{(A}\tilde{\FF}_{B)D}
     -  \frac{2}{r} (\bar{D}_{(A}r) \bar{D}_{B)}\tilde{F}
     \nonumber\\
  \!\!\!\!&=&\!\!\!\!
  16 \pi S_{(\FF)AB}
              ,
              \label{eq:1st-pert-Einstein-non-vac-AB-traceless-final}
  \\
  S_{(\FF)AB}
  \!\!\!\!&:=&\!\!\!\!
      T_{AB} - \frac{1}{2} y_{AB} T_{C}^{\;\;\;C}
      - 2 \left( \bar{D}_{(A}(r \tilde{T}_{(e1)B)}) - \frac{1}{2} y_{AB} \bar{D}^{D}(r \tilde{T}_{(e1)D}) \right)
      \nonumber\\
  &&\!\!\!\!
      + 2 \left( (\bar{D}_{(A}r) \bar{D}_{B)} - \frac{1}{2} y_{AB} (\bar{D}^{D}r) \bar{D}_{D}  \right) ( r \tilde{T}_{(e2)} )
      \nonumber\\
  &&\!\!\!\!
      + r \left( \bar{D}_{A}\bar{D}_{B} - \frac{1}{2} y_{AB} \bar{D}^{D}\bar{D}_{D} \right)(r \tilde{T}_{(e2)})
      \nonumber\\
  &&\!\!\!\!
      + 2 \left( (\bar{D}_{A}r) (\bar{D}_{B}r) - \frac{1}{2} y_{AB} (\bar{D}^{C}r) (\bar{D}_{C}r) \right) \tilde{T}_{(e2)}
      \nonumber\\
  &&\!\!\!\!
      + 2 y_{AB} (\bar{D}^{C}r) \tilde{T}_{(e1)C}
      -  r y_{AB} (\bar{D}^{C}r) \bar{D}_{C}\tilde{T}_{(e2)}
     .
  \label{eq:souce(FF)-def}
\end{eqnarray}


Equations (\ref{eq:linearized-Einstein-pq-traceless-even}),
(\ref{eq:even-FAB-divergence}),
(\ref{eq:even-mode-tildeF-master-eq}), and
(\ref{eq:1st-pert-Einstein-non-vac-AB-traceless-final}) are all
independent equations of the linearized Einstein equation for
even-mode perturbations.
These equations are coupled equations for the variables
$\tilde{F}_{C}^{\;\;\;C}$, $F$, and $\tilde{\FF}_{AB}$ and the
energy-momentum tensor for the matter field.
When we solve these equations, we have to take into account of the
continuity equations
(\ref{eq:div-barTab-linear-vac-back-u-A-mode}) and
(\ref{eq:div-barTab-linear-vac-back-u-p-mode-even}) for the
matter fields.
We note that these equations are valid not only for $l\geq 2$ modes
but also $l=0,1$ modes in our formulation.
For $l\geq 2$ modes, we can derive the Zerilli equation, while we can
derive formal solutions for $l=0,1$ modes.
The derivations of these formal solutions for $l=0,1$ modes are the
main ingredients of this paper.


\section{Component treatment of the even-mode linearized Einstein
  equations}
\label{sec:Schwarzschild_Background-non-vaccum-even-treatment}


To summarize the even-mode Einstein equations, we consider the static
chart of $y_{AB}$ as
Eq.~(\ref{eq:background-metric-2+2-y-comp-Schwarzschild}).
On this chart, the components of the Christoffel symbol
$\bar{\Gamma}_{AB}^{\;\;\;\;\;\;C}$ associated with the covariant
derivative $\bar{D}_{A}$ is summarized as
\begin{eqnarray}
  \bar{\Gamma}_{tt}^{\;\;\;\;t} = 0, \quad
  \bar{\Gamma}_{tr}^{\;\;\;\;t} = \frac{f'}{2f}, \quad
  \bar{\Gamma}_{rr}^{\;\;\;\;t} = 0, \quad
  \bar{\Gamma}_{tt}^{\;\;\;\;r} = \frac{f f'}{2}, \quad
  \bar{\Gamma}_{tr}^{\;\;\;\;r} = 0, \quad
  \bar{\Gamma}_{rr}^{\;\;\;\;r} = - \frac{f'}{2f},
\end{eqnarray}
where $f':=\partial_{r}f$.


First, Eq.~(\ref{eq:linearized-Einstein-pq-traceless-even}) is a
direct consequence of the even-mode Einstein equation.
Here, we introduce the components $X_{(e)}$ and $Y_{(e)}$ of the
traceless variable $\tilde{\FF}_{AB}$ by
\begin{eqnarray}
  \label{eq:Xe-Ye-def}
  \tilde{\FF}_{AB}
  =:
  X_{(e)} \left\{ - f (dt)_{A}(dt)_{B} - f^{-1} (dr)_{A}(dr)_{B}\right\}
  +
  2 Y_{(e)} (dt)_{(A}(dr)_{B)}
  .
\end{eqnarray}
Through these components $X_{(e)}$ and $Y_{(e)}$, $t$- and $r$-
components of Eq.~(\ref{eq:even-FAB-divergence}) are given by
\begin{eqnarray}
  &&
     \partial_{t}X_{(e)}
     + f \partial_{r}Y_{(e)}
     + f' Y_{(e)}
     - \frac{1}{2} \partial_{t}\tilde{F}
     =
     16 \pi S_{(ec)t}
     ,
     \label{eq:even-FAB-divergence-3-t-comp}
  \\
  &&
     \frac{1}{f} \partial_{t}Y_{(e)}
     + \partial_{r}X_{(e)}
     + \frac{f'}{f} X_{(e)}
     + \frac{1}{2} \partial_{r}\tilde{F}
     =
     - 16 \pi S_{(ec)r}
     .
     \label{eq:even-FAB-divergence-3-r-comp}
\end{eqnarray}
The source term $S_{(ec)A}$ is defined by
\begin{eqnarray}
  S_{(ec)A}
  :=
  r \tilde{T}_{(e1)A}
  -  \frac{1}{2} r^{2} \bar{D}_{A}\tilde{T}_{(e2)}
  .
  \label{eq:SecA-def}
\end{eqnarray}


Furthermore, the evolution equation
(\ref{eq:even-mode-tildeF-master-eq}) for the variable $\tilde{F}$ is
given by
\begin{eqnarray}
  -  \partial_{t}^{2}\tilde{F}
  + f \partial_{r}( f \partial_{r}\tilde{F} )
  + \frac{2}{r} f^{2} \partial_{r}\tilde{F}
  -  \frac{(l-1)(l+2)}{r^{2}} f \tilde{F}
  + \frac{4}{r^{2}} f^{2} X_{(e)}
  =
  16 \pi G f S_{(F)}
  .
  \label{eq:even-mode-tildeF-master-eq-mod-3-component}
\end{eqnarray}
The source term $S_{(F)}$ is defined by
\begin{eqnarray}
  S_{(F)}
  &:=&
       \tilde{T}_{E}^{\;\;\;E}
       + 4 (\bar{D}_{D}r) \tilde{T}_{(e1)}^{\;\;\;\;D}
       -  2 r (\bar{D}_{D}r)  \bar{D}^{D}\tilde{T}_{(e2)}
       -  (l(l+1)+2) \tilde{T}_{(e2)}
       \label{eq:source(F)-def-sum-2}
  \\
  &=&
      -  \frac{1}{f} \tilde{T}_{tt}
      + f \tilde{T}_{rr}
      + 4 f \tilde{T}_{(e1)r}
      -  2 r f \partial_{r}\tilde{T}_{(e2)}
      -  (l(l+1)+2) \tilde{T}_{(e2)}
      .
      \label{eq:source(F)-def-sum-component}
\end{eqnarray}


The component expression of
Eq.~(\ref{eq:1st-pert-Einstein-non-vac-AB-traceless-final}) with the
constraints (\ref{eq:even-FAB-divergence-3-t-comp}) and
(\ref{eq:even-FAB-divergence-3-r-comp}) are given by
\begin{eqnarray}
  &&
     \partial_{t}^{2}X_{(e)}
     - f \partial_{r}(f \partial_{r}X_{(e)})
     -  \frac{2(1-2f)f}{r} \partial_{r}X_{(e)}
     - \frac{(1-f)(1-5f)-l(l+1)f}{r^{2}} X_{e}
     \nonumber\\
  &&
     - \frac{(1-3f)f}{r} \partial_{r}\tilde{F}
     \nonumber\\
  &=&
      - 16 \pi
      \left(
      S_{(\FF)tt}
      + \frac{2f(3f-1)}{r} S_{(ec)r}
      \right)
      ,
      \label{eq:1st-pert-Ein-non-vac-tt-sum-2-2}
  \\
  &&
     \partial_{t}^{2}Y_{(e)}
     - f \partial_{r}\left( f \partial_{r}Y_{(e)} \right)
     - \frac{2(1-2f)f}{r} \partial_{r}Y_{(e)}
     - \frac{(1-f)(1-5f)-l(l+1)f}{r^{2}} Y_{(e)}
     \nonumber\\
  &&
     + \frac{1-3f}{r} \partial_{t}\tilde{F}
     \nonumber\\
  &=&
      16 \pi
      \left(
      f S_{(\FF)tr}
      - \frac{2(1-2f)}{r} S_{(ec)t}
      \right)
      .
      \label{eq:1st-pert-Ein-non-vac-tr-sum-2-2}
\end{eqnarray}
Here, we note that $(rr)$-component of
Eq.~(\ref{eq:1st-pert-Einstein-non-vac-AB-traceless-final}) with the
constraint (\ref{eq:even-FAB-divergence-3-r-comp}) is equivalent to
Eq.~(\ref{eq:1st-pert-Ein-non-vac-tt-sum-2-2}).
The source terms $S_{(\FF)tt}$ and $S_{(\FF)tr}$ in
Eqs.~(\ref{eq:1st-pert-Ein-non-vac-tt-sum-2-2}) and
(\ref{eq:1st-pert-Ein-non-vac-tr-sum-2-2}) are given by
\begin{eqnarray}
  S_{(\FF)tt}
  &=&
      \frac{1}{2} \left( \tilde{T}_{tt} + f^{2} \tilde{T}_{rr} \right)
      -    r          \partial_{t}\tilde{T}_{(e1)t}
      - 3 f^{2}   \tilde{T}_{(e1)r}
      -    r f^{2} \partial_{r}\tilde{T}_{(e1)r}
      \nonumber\\
  &&
     + \frac{r^{2}}{2}          \partial_{t}^{2}\tilde{T}_{(e2)}
     + \frac{r^{2}}{2} f^{2} \partial_{r}^{2}\tilde{T}_{(e2)}
     + 3 r f^{2}                       \partial_{r}\tilde{T}_{(e2)}
     + 2 f^{2}                         \tilde{T}_{(e2)}
     ,
     \label{eq:linear-even-source-tildeSAB-def-red-sum-tt-sum}
  \\
  S_{(\FF)tr}
  &=&
      \tilde{T}_{tr}
      -  r \partial_{t}\tilde{T}_{(e1)r}
      -  r \partial_{r}\tilde{T}_{(e1)t}
      -  \tilde{T}_{(e1)t}
      + \frac{1-f}{f} \tilde{T}_{(e1)t}
      \nonumber\\
  &&
     + r^{2} \partial_{t}\partial_{r}\tilde{T}_{(e2)}
     + 2r \partial_{t}\tilde{T}_{(e2)}
     - \frac{r(1-f)}{2f} \partial_{t}\tilde{T}_{(e2)}
     .
     \label{eq:linear-even-source-tildeSAB-def-red-sum-tr-sum}
\end{eqnarray}


The components of
Eq.~(\ref{eq:div-barTab-linear-vac-back-u-A-mode}) is given by
\begin{eqnarray}
  &&
     -  \partial_{t}\tilde{T}_{tt}
     + f^{2} \partial_{r}\tilde{T}_{rt}
     + \frac{(1+f)f}{r} \tilde{T}_{rt}
     -  \frac{f}{r} l(l+1) \tilde{T}_{(e1)t}
     =
     0
     ,
     \label{eq:div-barTab-linear-AB-t}
  \\
  &&
     - \partial_{t}\tilde{T}_{tr}
     + \frac{1-f}{2rf} \tilde{T}_{tt}
     + f^{2} \partial_{r}\tilde{T}_{rr}
     + \frac{(3+f)f}{2r} \tilde{T}_{rr}
     -  \frac{f}{r} l(l+1) \tilde{T}_{(e1)r}
     -  \frac{f}{r} \tilde{T}_{(e0)}
     =
     0
     ,
     \label{eq:div-barTab-linear-AB-r}
\end{eqnarray}
where Eq.~(\ref{eq:div-barTab-linear-AB-t}) is the $t$-component and
Eq.~(\ref{eq:div-barTab-linear-AB-r}) is the $r$-component,
respectively.
Furthermore, Eq.~(\ref{eq:div-barTab-linear-vac-back-u-p-mode-even})
is given by
\begin{eqnarray}
  - \partial_{t}\tilde{T}_{(e1)t}
  + f^{2} \partial_{r}\tilde{T}_{(e1)r}
  + \frac{(1+2f)f}{r} \tilde{T}_{(e1)r}
  + \frac{f}{2r} \tilde{T}_{(e0)}
  -  \frac{f}{2r} (l-1)(l+2) \tilde{T}_{(e2)}
  =
  0
  .
  \label{eq:div-barTab-linear-p-even-mode}
\end{eqnarray}


From the time derivative of
Eqs.~(\ref{eq:even-FAB-divergence-3-t-comp}) and
(\ref{eq:even-FAB-divergence-3-r-comp}), we obtain
\begin{eqnarray}
  &&
     \partial_{t}^{2}X_{(e)}
     -  f \partial_{r}(f \partial_{r}X_{(e)} )
     -  2 \frac{1-f}{r} f \partial_{r}X_{(e)}
     -  \frac{(1-3f)(1-f)}{r^{2}} X_{(e)}
     \nonumber\\
  && \quad\quad\quad
     - \frac{1}{2} \partial_{t}^{2}\tilde{F}
     - \frac{1}{2} f \partial_{r}(f \partial_{r}\tilde{F})
     - \frac{1-f}{2r} f \partial_{r}\tilde{F}
     \nonumber\\
  && \quad\quad\quad
     - 16 \pi \partial_{t}S_{(ec)t}
     - 16 \pi \partial_{r}(f^{2} S_{(ec)r})
     =
     0
     .
     \label{eq:even-FAB-divergence-3-t+r}
  \\
  &&
     -  \partial_{t}^{2}Y_{(e)}
     + f \partial_{r}( f \partial_{r}Y_{(e)} )
     + 2 f \frac{1-f}{r} \partial_{r}Y_{(e)}
     + \frac{(1-3f)(1-f)}{r^{2}} Y_{(e)}
     \nonumber\\
  && \quad\quad\quad
     -  \frac{1-f}{2r} \partial_{t}\tilde{F}
     -  f \partial_{r}\partial_{t}\tilde{F}
     -  16 \pi \partial_{r}( f S_{(ec)t} )
     -  16 \pi f \partial_{t}S_{(ec)r}
     =
     0
     .
     \label{eq:even-FAB-divergence-3-r-t}
\end{eqnarray}
From Eqs.~(\ref{eq:1st-pert-Ein-non-vac-tt-sum-2-2}) and
(\ref{eq:even-FAB-divergence-3-t+r}), we obtain
\begin{eqnarray}
  &&
     4 f \partial_{r}(f X_{(e)})
     + \frac{2}{r} l(l+1) (f X_{(e)})
     + r \partial_{t}^{2}\tilde{F}
     + r f \partial_{r}(f \partial_{r}\tilde{F})
     + (5f-1) f \partial_{r}\tilde{F}
     \nonumber\\
     &=&
         - 32 \pi r \left[
         S_{(\FF)tt}
         + \partial_{t}S_{(ec)t}
         + f^{2} \partial_{r}S_{(ec)r}
         + \frac{4}{r} f^{2} S_{(ec)r}
         \right]
         .
         \label{eq:1st-pert-Ein-non-vac-tt-FAB-div-t+r}
\end{eqnarray}
Furthermore, from Eqs.~(\ref{eq:1st-pert-Ein-non-vac-tr-sum-2-2}) and
(\ref{eq:even-FAB-divergence-3-r-t}), we obtain
\begin{eqnarray}
  &&
     4 f  \partial_{r}(f Y_{(e)})
     + \frac{2}{r} l(l+1) (f Y_{(e)})
     -  2 r f \partial_{r}\partial_{t}\tilde{F}
     -  (5f-1) \partial_{t}\tilde{F}
     \nonumber\\
  &=&
     32 \pi r \left[
     f S_{(\FF)tr}
     + f \partial_{t}S_{(ec)r}
     -  \frac{1-3f}{r} S_{(ec)t}
     + f \partial_{r}S_{(ec)t}
     \right]
      .
      \label{eq:1st-pert-Ein-non-vac-tr-FAB-div-t-r}
\end{eqnarray}
Equations~(\ref{eq:even-FAB-divergence-3-t-comp}) and
(\ref{eq:1st-pert-Ein-non-vac-tr-FAB-div-t-r}) yields
\begin{eqnarray}
  l(l+1) Y_{(e)}
  &=&
      r \partial_{t}\left(
      2 X_{(e)}
      + r \partial_{r}\tilde{F}
      \right)
      +  \frac{3f-1}{2f} r \partial_{t}\tilde{F}
      \nonumber\\
  &&
      + 16 \pi r^{2} \left[
      S_{(\FF)tr}
      + \partial_{t}S_{(ec)r}
      -  \frac{1-f}{rf} S_{(ec)t}
      + \partial_{r}S_{(ec)t}
      \right]
     .
      \label{eq:ll+1fYe-1st-pert-Ein-non-vac-tr-FAB-div-t-r}
\end{eqnarray}
Similarly, Equations~(\ref{eq:1st-pert-Ein-non-vac-tt-FAB-div-t+r})
and (\ref{eq:even-mode-tildeF-master-eq-mod-3-component}) yield
\begin{eqnarray}
  &&
     4 f \partial_{r}(f X_{(e)})
     + \frac{2}{r} [ l(l+1) + 2f ] (f X_{(e)})
     \nonumber\\
     &&
     + 2 f \partial_{r}( r f \partial_{r}\tilde{F} )
     + (5f-1) f \partial_{r}\tilde{F}
     -  \frac{(l-1)(l+2)}{r} f \tilde{F}
     \nonumber\\
     &=&
         -  32 \pi r \left[
         S_{(\FF)tt}
         + \partial_{t}S_{(ec)t}
         + f^{2} \partial_{r}S_{(ec)r}
         + \frac{4}{r} f^{2} S_{(ec)r}
         -  \frac{f}{2} S_{(F)}
         \right]
         .
         \label{eq:1st-pert-Ein-non-vac-tt-FAB-div-t+r-constraint}
\end{eqnarray}


Thus, we may regard that the independent components of the Einstein
equations for the even-mode perturbations are summarized as
Eqs.~(\ref{eq:even-mode-tildeF-master-eq-mod-3-component}),
(\ref{eq:1st-pert-Ein-non-vac-tt-sum-2-2}),
(\ref{eq:ll+1fYe-1st-pert-Ein-non-vac-tr-FAB-div-t-r}), and
(\ref{eq:1st-pert-Ein-non-vac-tt-FAB-div-t+r-constraint}).


As shown in many
literatures~\cite{F.Zerilli-1970-PRL,F.Zerilli-1970-PRD,H.Nakano-2019,V.Moncrief-1974a,V.Moncrief-1974b,C.T.Cunningham-R.H.Price-V.Moncrief-1978},
it is well-known that
Eqs.~(\ref{eq:even-mode-tildeF-master-eq-mod-3-component}),
(\ref{eq:1st-pert-Ein-non-vac-tt-sum-2-2}), and
(\ref{eq:1st-pert-Ein-non-vac-tt-FAB-div-t+r-constraint}) are reduced
to the single master equation for a single variable.
We trace this procedure.


Equation~(\ref{eq:1st-pert-Ein-non-vac-tt-FAB-div-t+r-constraint}) is an
initial value constraint for the variables $(X_{(e)},\tilde{F})$, while
Eqs.~(\ref{eq:even-mode-tildeF-master-eq-mod-3-component}) and
(\ref{eq:even-FAB-divergence-3-t+r}) are evolution equations.
Equation (\ref{eq:ll+1fYe-1st-pert-Ein-non-vac-tr-FAB-div-t-r})
directly yields that the variable $Y_{(e)}$ is determined by the
solution $(X_{(e)},\tilde{F})$ to
Eqs.~(\ref{eq:1st-pert-Ein-non-vac-tt-FAB-div-t+r-constraint}),
(\ref{eq:even-mode-tildeF-master-eq-mod-3-component}) and
(\ref{eq:even-FAB-divergence-3-t+r}), if $l\neq 0$.
If the initial value constraint
(\ref{eq:1st-pert-Ein-non-vac-tt-FAB-div-t+r-constraint}) is
reduced to the equation of a variable $\Phi_{(e)}$ and
$\tilde{F}$, we may expect that $\Phi_{(e)}$ linearly depends on
$f X_{(e)}$, $\tilde{F}$, and $r f \partial_{r}\tilde{F}$.
To show this, we introduce the variable $\Phi_{(e)}$ as
\begin{eqnarray}
  \label{eq:Moncrief-master-variable-tryal}
  \alpha \Phi_{(e)} := f X_{(e)} + \beta \tilde{F} + \gamma r f \partial_{r}\tilde{F},
\end{eqnarray}
where $\alpha$, $\beta$, and $\gamma$ may depend on $r$.
Substituting Eq.~(\ref{eq:Moncrief-master-variable-tryal}) into Eq.~(\ref{eq:1st-pert-Ein-non-vac-tt-FAB-div-t+r-constraint}), we obtain
\begin{eqnarray}
  0
  &=&
      -  4 r f \alpha' \Phi_{(e)}
      -  2 [ l(l+1) + 2f ] \alpha \Phi_{(e)}
      -  4 r f \alpha \partial_{r}\Phi_{(e)}
      + 4 \left( \gamma - \frac{1}{2} \right) r f \partial_{r}[ r f \partial_{r}\tilde{F} ]
      \nonumber\\
  &&
     +
     \left[
     4 \beta
     + 4 r f \gamma'
     + 2 \left\{ l(l+1) + 2f \right\} \gamma
     -  (5f-1)
     \right] r f \partial_{r}\tilde{F}
     \nonumber\\
  &&
     +
     \left[
     4 r f \beta'
     + 2 \left\{ l(l+1) + 2f \right\} \beta
     +  (l-1)(l+2) f
     \right] \tilde{F}
     \nonumber\\
  &&
     - 32 \pi r^{2} \left[
     S_{(\FF)tt}
     + \partial_{t}S_{(ec)t}
     + f^{2} \partial_{r}S_{(ec)r}
     + \frac{4}{r} f^{2} S_{(ec)r}
     -  \frac{f}{2} S_{(F)}
     \right]
     .
     \label{eq:1st-pert-Ein-non-vac-tt-FAB-div-t+r-constraint-intermed-1}
\end{eqnarray}
Here, we choose
\begin{eqnarray}
  \gamma = \frac{1}{2}
\end{eqnarray}
to eliminate the term of the second derivative of $\tilde{F}$.
Owing to this choice, we obtain
\begin{eqnarray}
  0
  &=&
      -  4 r f \alpha' \Phi_{(e)}
      -  2 [ l(l+1) + 2f ] \alpha \Phi_{(e)}
      -  4 r f \alpha \partial_{r}\Phi_{(e)}
      + [ 4 \beta + \Lambda ] r f \partial_{r}\tilde{F}
      \nonumber\\
  &&
     + \left[
     4 r f \beta'
     + 2 \left\{ l(l+1) + 2f \right\} \beta
     +  (l-1)(l+2) f
     \right] \tilde{F}
     \nonumber\\
  &&
     - 32 \pi r^{2} \left[
     S_{(\FF)tt}
     + \partial_{t}S_{(ec)t}
     + f^{2} \partial_{r}S_{(ec)r}
     + \frac{4}{r} f^{2} S_{(ec)r}
     -  \frac{f}{2} S_{(F)}
     \right]
     .
     \label{eq:1st-pert-Ein-non-vac-tt-FAB-div-t+r-constraint-intermed-2}
\end{eqnarray}
Here, we choose $\beta$ as
\begin{eqnarray}
  \beta = - \frac{1}{4} \Lambda := - \frac{1}{4} [ (l-1)(l+2) + 3(1-f) ], \quad
  \Lambda := (l-1)(l+2) + 3(1-f)
\end{eqnarray}
to eliminate the term of the first derivative of $\tilde{F}$.
Due to this choice, we obtain
\begin{eqnarray}
  l(l+1) \Lambda \tilde{F}
  &=&
      -  8 r f \partial_{r}\left(\alpha \Phi_{(e)}\right)
      -  4 [ l(l+1) + 2f ] \alpha \Phi_{(e)}
      \nonumber\\
  &&
     - 64 \pi r^{2} \left[
     S_{(\FF)tt}
     + \partial_{t}S_{(ec)t}
     + f^{2} \partial_{r}S_{(ec)r}
     + \frac{4}{r} f^{2} S_{(ec)r}
     -  \frac{f}{2} S_{(F)}
     \right]
     .
     \label{eq:1st-pert-Ein-non-vac-XF-constraint-apha-Phie}
\end{eqnarray}
This equation yields that the variable $\tilde{F}$ is determined by
the single variable $\Phi_{(e)}$ and the source terms if $l\neq 0$ and
if the coefficient $\alpha$ is determined.


At this moment, the variable $\Phi_{(e)}$ is determined up to its
normalization $\alpha$ as
\begin{eqnarray}
  \label{eq:Moncrief-master-variable-intermediate}
  \alpha \Phi_{(e)} := f X_{(e)} - \frac{1}{4} \Lambda \tilde{F} + \frac{1}{2} r f \partial_{r}\tilde{F}.
\end{eqnarray}
Eliminating $X_{(e)}$ in
Eq.~(\ref{eq:even-mode-tildeF-master-eq-mod-3-component}) through
Eq.~(\ref{eq:Moncrief-master-variable-intermediate}), we obtain
\begin{eqnarray}
  -  \partial_{t}^{2}\tilde{F}
  + f \partial_{r}( f \partial_{r}\tilde{F} )
  + \frac{1}{r^{2}} 3(1-f) f \tilde{F}
  + \frac{4}{r^{2}} f \alpha \Phi_{(e)}
  =
  16 \pi f S_{(F)}
  .
  \label{eq:even-mode-tildeF-eq-Phie}
\end{eqnarray}
Similarly, eliminating $X_{(e)}$ in
Eq.~(\ref{eq:1st-pert-Ein-non-vac-tt-sum-2-2}) through
Eqs.~(\ref{eq:1st-pert-Ein-non-vac-XF-constraint-apha-Phie})--(\ref{eq:even-mode-tildeF-eq-Phie}),
we obtain
\begin{eqnarray}
  0
  &=&
      -  \alpha \partial_{t}^{2}\Phi_{(e)}
      + \alpha f \partial_{r}\left[ f \partial_{r}\Phi_{(e)} \right]
      + 2 \alpha \left[
      \frac{\alpha'}{\alpha}
      + \frac{1}{r}
      + \frac{1}{r\Lambda} 3 (1-f)
      \right] f^{2} \partial_{r}\Phi_{(e)}
     \nonumber\\
  &&
     + \left[
     \alpha'' f
     + \alpha' \frac{1}{r} (1+f)
     + \alpha' \frac{1}{r\Lambda} 3(1-f) 2f
     + \alpha  \frac{3(1-f)\left\{l(l+1)+2f\right\}}{r^{2}\Lambda}
     \right.
      \nonumber\\
  && \quad\quad
     \left.
     - \alpha \frac{(l-1)(l+2)+1+f}{r^{2}}
     \right] f \Phi_{(e)}
      \nonumber\\
  &&
      + 16 \pi f \frac{\Lambda + 3(1-f)}{\Lambda} \left[
      S_{(\FF)tt}
      + \partial_{t}S_{(ec)t}
      + f^{2} \partial_{r}S_{(ec)r}
      + \frac{4f^{2}}{r} S_{(ec)r}
      -  \frac{f}{2} S_{(F)}
      \right]
      \nonumber\\
  &&
      -  16 \pi f S_{(\FF)tt}
      -  32 \pi \frac{3f-1}{r} f^{2} S_{(ec)r}
      -  16 \pi ( -  \frac{1}{4} \Lambda ) f S_{(F)}
     - 16 \pi r \frac{1}{2} f \partial_{r}\left[ f S_{(F)} \right]
     .
     \label{eq:1st-pert-Ein-non-vac-tt-2-3-2-with-Phie-pre}
\end{eqnarray}
We determine $\alpha$ so that the terms proportional to
$\partial_{r}\Phi_{(e)}$ vanish.
Then, we obtain the equation for $\alpha$ as
\begin{eqnarray}
  \frac{\alpha'}{\alpha}
  + \frac{1}{r}
  + \frac{1}{r\Lambda} 3(1-f)
  =
  0
  .
\end{eqnarray}
From this equation, we obtain
\begin{eqnarray}
  \label{eq:Moncrief-master-alpha-sol}
  \frac{1}{\alpha}
  =
  \frac{Cr}{\Lambda},
\end{eqnarray}
where $C$ is a constant of integration.
In this paper, we choose $C=1$.
Then, we obtain
\begin{eqnarray}
  \label{eq:Moncrief-master-variable-final}
  \Phi_{(e)}
  :=
  \frac{1}{\alpha} \left[
  f X_{(e)}
  - \frac{1}{4} \Lambda \tilde{F}
  + \frac{1}{2} r f \partial_{r}\tilde{F}
  \right]
  =
  \frac{r}{\Lambda} \left[
  f X_{(e)}
  - \frac{1}{4} \Lambda \tilde{F}
  + \frac{1}{2} r f \partial_{r}\tilde{F}
  \right]
  .
\end{eqnarray}
This is the Moncrief variable.


From Eq.~(\ref{eq:Moncrief-master-alpha-sol}), we obtain
\begin{eqnarray}
  \alpha'
  =
  - \frac{1}{r} \alpha
  - \frac{1}{\Lambda} 3 \frac{1-f}{r} \alpha
  , \quad
  \alpha''
  =
  + \frac{2}{r^{2}} \alpha
  + \frac{12(1-f)}{\Lambda r^{2}} \alpha
  .
\end{eqnarray}
Then, using
\begin{eqnarray}
  \mu := (l-1)(l+2), \quad \Lambda = \mu + 3(1-f),
\end{eqnarray}
Eq.~(\ref{eq:1st-pert-Ein-non-vac-tt-2-3-2-with-Phie-pre}) is given by
\begin{eqnarray}
  &&
     -  \frac{1}{f} \partial_{t}^{2}\Phi_{(e)}
     + \partial_{r}\left[ f \partial_{r}\Phi_{(e)} \right]
     -
     \frac{1}{r^{2}\Lambda^{2}}
     \left[
     \mu^{2} [ (\mu+2) + 3(1-f) ]
     +
     9(1-f)^{2}
     \left(
     \mu
     + 1-f
     \right)
     \right] \Phi_{(e)}
     \nonumber\\
  &=&
      16 \pi \frac{r}{\Lambda} \left[
      -  \partial_{t}S_{(ec)t}
      -  f^{2} \partial_{r}S_{(ec)r}
      + 2f \frac{f-1}{r} S_{(ec)r}
     + \frac{r}{2} f \partial_{r}S_{(F)}
     + \frac{1}{2} S_{(F)}
     -  \frac{1}{4} \Lambda S_{(F)}
     \right.
     \nonumber\\
  && \quad\quad\quad\quad
     \left.
     + \frac{3(1-f)}{\Lambda} \left[
     -  S_{(\FF)tt}
     -  \partial_{t}S_{(ec)t}
     -  f^{2} \partial_{r}S_{(ec)r}
     -  \frac{4}{r} f^{2} S_{(ec)r}
     + \frac{f}{2} S_{(F)}
     \right]
     \right]
     .
     \label{eq:Zerilli-Moncrief-eq-final}
\end{eqnarray}
This is the Zerilli equation for the Moncrief variable
(\ref{eq:Moncrief-master-variable-final}).


Here, we summarize the equations for even-mode perturbations.
We derive the definition of the Moncrief variable as
Eq.~(\ref{eq:Moncrief-master-variable-final}), i.e.,
\begin{eqnarray}
  \label{eq:Moncrief-master-variable-final-sum}
  \Phi_{(e)}
  :=
  \frac{r}{\Lambda} \left[
  f X_{(e)}
  - \frac{1}{4} \Lambda \tilde{F}
  + \frac{1}{2} r f \partial_{r}\tilde{F}
  \right]
  ,
\end{eqnarray}
where $\Lambda$ is defined by
\begin{eqnarray}
  \label{eq:mu-Lambda-defs}
  \Lambda = \mu + 3(1-f), \quad \mu := (l-1)(l+2).
\end{eqnarray}
This definition of the variable $\Phi_{(e)}$ implies that if we obtain
the variables $\Phi_{(e)}$ and $\tilde{F}$ are determined, the
component of $X_{(e)}$ of the metric perturbation is determined
through the equation
\begin{eqnarray}
  f X_{(e)}
  =
  \frac{1}{r} \Lambda \Phi_{(e)}
  + \frac{1}{4} \Lambda \tilde{F}
  - \frac{1}{2} r f \partial_{r}\tilde{F}
  .
  \label{eq:Moncrief-master-variable-final-sum-fX=}
\end{eqnarray}
As the initial value constraint for the variable $\tilde{F}$ and
$Y_{(e)}$, we have
Eqs.~(\ref{eq:1st-pert-Ein-non-vac-XF-constraint-apha-Phie}) and
(\ref{eq:ll+1fYe-1st-pert-Ein-non-vac-tr-FAB-div-t-r}) as
\begin{eqnarray}
  l(l+1) \Lambda \tilde{F}
  &=&
      -  8 f \Lambda \partial_{r}\Phi_{(e)}
      + \frac{4}{r} \left[ 6 f (1-f) - l(l+1) \Lambda \right] \Phi_{(e)}
      - 64 \pi r^{2} S_{(\Lambda\tilde{F})}
      ,
      \label{eq:-Ein-non-vac-XF-constraint-apha-Phie-with-SLambdaF-sum}
  \\
  l(l+1) Y_{(e)}
  &=&
      r \partial_{t}\left(
      2 X_{(e)}
      + r \partial_{r}\tilde{F}
      \right)
      +  \frac{3f-1}{2f} r \partial_{t}\tilde{F}
      + 16 \pi r^{2} S_{(Y_{(e)})}
      ,
      \label{eq:ll+1fYe-1st-pert-Ein-non-vac-tr-FAB-div-t-r-sum-3}
\end{eqnarray}
where the source term $S_{(\Lambda\tilde{F})}$ and $S_{(Y_{(e)})}$ are
given by
\begin{eqnarray}
  S_{(\Lambda\tilde{F})}
  &:=&
       S_{(\FF)tt}
       + \partial_{t}S_{(ec)t}
       + f^{2} \partial_{r}S_{(ec)r}
       + \frac{4}{r} f^{2} S_{(ec)r}
       -  \frac{f}{2} S_{(F)}
       \label{eq:SLambdaF-def}
  \\
  &=&
      \tilde{T}_{tt}
      + r f^{2} \partial_{r}\tilde{T}_{(e2)}
      + 2 f (f+1) \tilde{T}_{(e2)}
      +  \frac{1}{2} f (l-1)(l+2) \tilde{T}_{(e2)}
      ,
      \label{eq:SLambdaF-def-explicit}
\end{eqnarray}
and
\begin{eqnarray}
  S_{(Y_{(e)})}
  &:=&
       S_{(\FF)tr}
       + \partial_{t}S_{(ec)r}
       -  \frac{1-f}{rf} S_{(ec)t}
       + \partial_{r}S_{(ec)t}
       \label{eq:sourceYe-def}
  \\
  &=&
      \tilde{T}_{tr}
      + r \partial_{t}\tilde{T}_{(e2)}
      .
      \label{eq:sourceYe-SPsie-explicit-sum}
\end{eqnarray}
Equation
(\ref{eq:-Ein-non-vac-XF-constraint-apha-Phie-with-SLambdaF-sum})
implies that the variable $\tilde{F}$ of the metric perturbation is
determined if the variable $\Phi_{(e)}$ and
source term $S_{(\Lambda\tilde{F})}$ are specified.
Equation (\ref{eq:ll+1fYe-1st-pert-Ein-non-vac-tr-FAB-div-t-r-sum-3})
implies that the component $Y_{(e)}$ of the metric perturbation is
determined if the variables $X_{(e)}$, $\tilde{F}$, and the source
term $S_{(Y_{(e)})}$ are specified.


Thus, apart from the source terms, the component $\tilde{F}$ of the
metric perturbation is determined through
Eq.~(\ref{eq:-Ein-non-vac-XF-constraint-apha-Phie-with-SLambdaF-sum})
if the Moncrief variable $\Phi_{(e)}$ is specified.
The component $X_{(e)}$ of the metric perturbation is determined
through Eq.~(\ref{eq:Moncrief-master-variable-final-sum-fX=}) if the
variables $\Phi_{(e)}$ and $\tilde{F}$ are specified.
Finally, the component $Y_{(e)}$ of the metric perturbation is
determined through
Eq.~(\ref{eq:ll+1fYe-1st-pert-Ein-non-vac-tr-FAB-div-t-r-sum-3}) if
the variables $\tilde{F}$ and $X_{(e)}$ are specified.
Namely, the components $X_{(e)}$, $Y_{(e)}$, and $\tilde{F}$ of the
metric perturbation are determined by the Moncrief variable
$\Phi_{(e)}$.
The Moncrief variable $\Phi_{(e)}$ is determined by the master
equation
\begin{eqnarray}
  -  \frac{1}{f} \partial_{t}^{2}\Phi_{(e)}
  + \partial_{r}\left[ f \partial_{r}\Phi_{(e)} \right]
  -
  V_{even} \Phi_{(e)}
  =
  16 \pi \frac{r}{\Lambda} S_{(\Phi_{(e)})}
  ,
  \label{eq:Zerilli-Moncrief-eq-final-sum}
\end{eqnarray}
where the potential function $V_{even}$ is defined by
\begin{eqnarray}
  V_{even}
  &:=&
      \frac{1}{r^{2}\Lambda^{2}}
      \left[
      \mu^{2} [ (\mu+2) + 3(1-f) ]
      +
      (3(1-f))^{2}
      \left(
      + \mu
      + (1-f)
      \right)
      \right]
      \nonumber\\
  &=&
      \frac{1}{r^{2}\Lambda^{2}}
      \left[
      \Lambda^{3}
      - 2 (2-3f) \Lambda^{2}
      + 6 (1-3f) (1-f) \Lambda
      + 18 f (1-f)^{2}
      \right]
      ,
      \label{eq:Zerilli-Moncrief-master-potential-final-sum}
\end{eqnarray}
and the source term in Eq.~(\ref{eq:Zerilli-Moncrief-eq-final-sum}) is
given by
\begin{eqnarray}
  S_{(\Phi_{(e)})}
  &:=&
       -  \partial_{t}S_{(ec)t}
       -  f^{2} \partial_{r}S_{(ec)r}
       + 2f \frac{f-1}{r} S_{(ec)r}
       + \frac{r}{2} f \partial_{r}S_{(F)}
     + \frac{1}{2} S_{(F)}
     -  \frac{1}{4} \Lambda S_{(F)}
     \nonumber\\
  &&
     + \frac{3(1-f)}{\Lambda} \left[
     -  S_{(\FF)tt}
     -  \partial_{t}S_{(ec)t}
     -  f^{2} \partial_{r}S_{(ec)r}
     -  \frac{4}{r} f^{2} S_{(ec)r}
     + \frac{f}{2} S_{(F)}
     \right]
     \label{eq:Zerilli-Moncrief-eq-source-def}
  \\
  &=&
      \frac{1}{2} \left(
      \frac{\Lambda}{2f} - 1
      \right) \tilde{T}_{tt}
      + \frac{1}{2} \left(
      (2-f) -  \frac{1}{2} \Lambda
      \right) f \tilde{T}_{rr}
      -  \frac{1}{2} r \partial_{r}\tilde{T}_{tt}
      + \frac{1}{2} f^{2} r \partial_{r}\tilde{T}_{rr}
      \nonumber\\
  &&
      -  \frac{f}{2} \tilde{T}_{(e0)}
      -  l(l+1) f \tilde{T}_{(e1)r}
      \nonumber\\
  &&
      + \frac{1}{2} r^{2} \partial_{t}^{2}\tilde{T}_{(e2)}
      -  \frac{1}{2} f^{2} r^{2} \partial_{r}^{2}\tilde{T}_{(e2)}
      - \frac{1}{2} 3(1+f) r f \partial_{r}\tilde{T}_{(e2)}
      \nonumber\\
  &&
      -  \frac{1}{2} (7-3f) f \tilde{T}_{(e2)}
      + \frac{1}{4} (l(l+1)-1-f) (l(l+1)+2) \tilde{T}_{(e2)}
      \nonumber\\
  &&
      - \frac{3(1-f)}{\Lambda} \left[
      \tilde{T}_{tt}
      + r f^{2} \partial_{r}\tilde{T}_{(e2)}
      + \frac{1}{2} (1+7f) f \tilde{T}_{(e2)}
      \right]
     .
      \label{eq:SPhie-def-explicit-sum}
\end{eqnarray}
To solve the master equation (\ref{eq:Zerilli-Moncrief-eq-final-sum})
we have to impose appropriate boundary conditions and solve as the
Cauchy problem.
In the book~\cite{Chandrasekhar-1983}, it is shown that the Zerilli
equation (\ref{eq:Zerilli-Moncrief-eq-final-sum}) without the source
term, i.e., $S_{(\Phi_{(e)})}=0$, can be transformed to the
Regge-Wheeler equation.
This transformation is called the Chandrasekhar transformation.
Since the Regge-Wheeler equation can be solved by MST (Mano Suzuki
Takasugi)
formulation~\cite{S.Mano-H.Suzuki-E.Takasugi-1996a,S.Mano-H.Suzuki-E.Takasugi-1996b,S.Mano-E.Takasugi-1997},
we may say that the solution to the Zerilli equation
(\ref{eq:Zerilli-Moncrief-eq-final-sum}) without the source term is
obtained through MST formulation.


Finally, we note that the solutions $\Phi_{(e)}$ and $\tilde{F}$ satisfy
the equation (\ref{eq:even-mode-tildeF-eq-Phie}),
as the consistency of the linearized Einstein equation.
Here, the source term $S_{(F)}$ is explicitly given by
Eq.~(\ref{eq:source(F)-def-sum-component}).
Here, we check this consistency of the initial value constraint
(\ref{eq:-Ein-non-vac-XF-constraint-apha-Phie-with-SLambdaF-sum}) and
the evolution equation (\ref{eq:even-mode-tildeF-eq-Phie}).
From Eqs.~(\ref{eq:even-mode-tildeF-eq-Phie}) and
(\ref{eq:Zerilli-Moncrief-eq-final-sum}), we obtain
\begin{eqnarray}
  0
  &=&
      r^{2} \Lambda \partial_{t}^{2}S_{(\Lambda\tilde{F})}
      -  \left[
      (5-3f) \Lambda
      + 3 (1-f) (1+f)
      + 18 \frac{1}{\Lambda} f (1-f)^{2}
      \right] f S_{(\Lambda\tilde{F})}
      \nonumber\\
  &&
     - 2 \left[
      3 (1-f) + 2 \Lambda
     \right] f^{2} r \partial_{r}S_{(\Lambda\tilde{F})}
      - \Lambda r^{2} f \partial_{r}\left[ f \partial_{r}S_{(\Lambda\tilde{F})} \right]
      \nonumber\\
  &&
     + \frac{1}{4}  \left[ (1-3f) - \Lambda  \right] \Lambda^{2} f S_{(F)}
      \nonumber\\
  &&
      -  2 r f^{2} \Lambda \partial_{r}S_{(\Phi_{(e)})}
     -  \left[ \Lambda + (1+3f) \right] \Lambda f S_{(\Phi_{(e)})}
      .
      \label{eq:even-mode-tildeF-eq-Phie-remainig-source-tmp}
\end{eqnarray}
This is an identity of the source terms.
We have confirmed
Eq.~(\ref{eq:even-mode-tildeF-eq-Phie-remainig-source-tmp}) is an
identity due to the definitions
(\ref{eq:SLambdaF-def-explicit})--(\ref{eq:source(F)-def-sum-component})
and the continuity equations
(\ref{eq:div-barTab-linear-AB-t})--(\ref{eq:div-barTab-linear-p-even-mode})
of the perturbative energy-momentum tensor.
This means that the evolution equation
(\ref{eq:even-mode-tildeF-eq-Phie}) is trivial when $l\neq 0$.
Thus, we have confirmed that the above strategy for $l\neq 0$ modes
are consistent.


Of course, this strategy is valid only when $l\neq 0$.
In the $l=0$ case, we have to consider the different strategy to
obtain the variable $X_{(e)}$, $Y_{(e)}$, and $\tilde{F}$.
This will be discussed Sec.~\ref{sec:l=0_Schwarzschild_Background-non-vac}.


Before going to the discussion on the strategy to solve $l=0$ mode
Einstein equation, we comment on the original equation derived by
Zerilli~\cite{F.Zerilli-1970-PRL,F.Zerilli-1970-PRD} for $l\geq 2$.
We consider the original time derivative of the Moncrief master
variable (\ref{eq:Moncrief-master-variable-final-sum}) as
\begin{eqnarray}
  \partial_{t}\Phi_{(e)}
  &=&
      \frac{r}{\Lambda} \left[
      f \partial_{t}X_{(e)}
      - \frac{1}{4} \Lambda \partial_{t}\tilde{F}
      + \frac{1}{2} r f \partial_{t}\partial_{r}\tilde{F}
      \right]
      .
      \label{eq:time-derivative-of-Moncrief-master-variable}
\end{eqnarray}
On the other hand,
Eq.~(\ref{eq:ll+1fYe-1st-pert-Ein-non-vac-tr-FAB-div-t-r-sum-3}) is
given by
\begin{eqnarray}
  \partial_{t}X_{(e)}
  &=&
      \frac{l(l+1)}{2r} Y_{(e)}
      - \frac{r}{2} \partial_{t}\partial_{r}\tilde{F}
      -  \frac{3f-1}{4f} \partial_{t}\tilde{F}
      - 8 \pi r S_{(Y_{(e)})}
      .
      \label{eq:ll+1fYe-1st-pert-Ein-non-vac-tr-FAB-div-t-r-4}
\end{eqnarray}
Substituting
Eq.~(\ref{eq:ll+1fYe-1st-pert-Ein-non-vac-tr-FAB-div-t-r-4}) into
Eq.~(\ref{eq:time-derivative-of-Moncrief-master-variable}), for
$l\neq 0$ modes, we obtain
\begin{eqnarray}
  \frac{1}{l(l+1)} \partial_{t}\Phi_{(e)}
  &=&
      \frac{1}{2\Lambda} \left[
      f Y_{(e)}
      -  \frac{r}{2} \partial_{t}\tilde{F}
      \right]
      - 8 \pi r^{2} f \frac{1}{l(l+1)\Lambda} S_{(Y_{(e)})}
      .
      \label{eq:time-derivative-of-Moncrief-master-variable-2}
\end{eqnarray}
Here, if we define the variable $\Psi_{(e)}$ by
\begin{eqnarray}
  \Psi_{(e)}
  &:=&
       \frac{1}{2\Lambda} \left[
       f Y_{(e)}
       -  \frac{r}{2} \partial_{t}\tilde{F}
       \right]
       \label{eq:Psie-def}
  \\
  &=&
      \frac{1}{l(l+1)} \partial_{t}\Phi_{(e)}
      + 8 \pi r^{2} f \frac{1}{l(l+1)\Lambda} S_{(Y_{(e)})}
      ,
      \label{eq:Zerilli-master-variable}
\end{eqnarray}
the variable $\Psi_{(e)}$ corresponds to original Zerilli's master
variable.
Roughly speaking, the variable $\Psi_{(e)}$ corresponds to the
time-derivative of the variable $\Phi_{(e)}$ with additional source
terms from the matter fields.
Therefore, it is trivial $\Psi_{(e)}$ also satisfies the Zerilli
equation with different source terms.
In other words, the Zerilli equation for $\Psi_{(e)}$ is derived by
the time derivative of the Zerilli equation for $\Phi_{(e)}$.
This means that the solution to the Zerilli equation for $\Psi_{(e)}$
may include an additional arbitrary function of $r$ as an
``integration constants.''
This ``integration constants'' do not included in the solution
$\Phi_{(e)}$ for the Zerilli equation
(\ref{eq:Zerilli-Moncrief-eq-final-sum}).
In this sense, the restriction of the initial value of
Eq.~(\ref{eq:Zerilli-Moncrief-eq-final-sum}) for $\Phi_{(e)}$ is
stronger than that of Eq.~(\ref{eq:Zerilli-Moncrief-eq-final-sum}) for
$\Psi_{(e)}$.


\section{$l=0$ mode perturbations on the Schwarzschild Background}
\label{sec:l=0_Schwarzschild_Background-non-vac}


Here, we consider the $l=0$ mode perturbations based on the
perturbation equations for the even-mode on Schwarzschild background
which are summarized in
Sec.~\ref{sec:Schwarzschild_Background-non-vaccum-even-treatment}.
Since
Proposal~\ref{proposal:treatment-proposal-on-pert-on-spherical-BG}
enable us to carry out the mode-by-mode analyses including $l=0,1$
modes, all equations in
Sec.~\ref{sec:Schwarzschild_Background-non-vaccum-even-treatment}
except for
Eqs.~(\ref{eq:time-derivative-of-Moncrief-master-variable-2}) and
(\ref{eq:Zerilli-master-variable}) are valid even in $l=0$ mode.
However, the strategy to solve these equations is different from that
$l\neq 0$ modes, because
Eqs.~(\ref{eq:-Ein-non-vac-XF-constraint-apha-Phie-with-SLambdaF-sum})
and (\ref{eq:ll+1fYe-1st-pert-Ein-non-vac-tr-FAB-div-t-r-sum-3}) do
not directly give the components $(\tilde{F},Y_{(e)})$ of the metric
perturbation for $l=0$ mode.


Before showing the strategy to solve even-mode Einstein equations for
$l=0$ mode, we note that
\begin{eqnarray}
  \label{eq:l=0DpkDelta-DpDqkDelta}
  \hat{D}_{p}k_{(\hat{\Delta})} = 0 = \hat{D}_{p}\hat{D}_{q}k_{(\hat{\Delta})}
\end{eqnarray}
if we impose the regularity $\delta=0$ to the harmonic function
$k_{(\hat{\Delta})}$.
In this case, the only remaining components of the linearized
energy-momentum tensor is
\begin{eqnarray}
  \label{eq:l=0-regularized-energy-momentum-tensor}
  T_{ab}
  =
  \tilde{T}_{AB} k_{(\hat{\Delta})} (dx^{A})_{a} (dx^{B})_{b}
  +
  \frac{1}{2} \gamma_{pq} \tilde{T}_{(e0)} k_{(\hat{\Delta})} (dx^{p})_{a} (dx^{q})_{b}.
\end{eqnarray}
Therefore, we can safely regard that
\begin{eqnarray}
  \label{eq:l=0-regularized-energy-momentum-tensor-equivalents}
  \tilde{T}_{(e2)}=0, \quad \tilde{T}_{(e1)A}=0.
\end{eqnarray}


Owing to
Eq.~(\ref{eq:l=0-regularized-energy-momentum-tensor-equivalents}), the
trace of the perturbation $\tilde{F}_{AB}$ is determined by the
Einstein equation (\ref{eq:linearized-Einstein-pq-traceless-even}), i.e.,
\begin{eqnarray}
  \tilde{F}_{D}^{\;\;\;D} = 0.
  \label{eq:eventildeFtrace-matter-reduce-2-l=0}
\end{eqnarray}
In the case of $l=0$ mode, $\Lambda$ defined by
Eq.~(\ref{eq:mu-Lambda-defs}) is given by
\begin{eqnarray}
  \label{eq:Lambda-def-l=0}
  \Lambda = 1-3f.
\end{eqnarray}
Then, the Moncrief master variable $\Phi_{(e)}$ is given by
Eq.~(\ref{eq:Moncrief-master-variable-final-sum}), i.e.,
\begin{eqnarray}
  \Phi_{(e)}
  :=
  \frac{r}{1-3f} \left[
  f X_{(e)}
  - \frac{1}{4} (1-3f) \tilde{F}
  + \frac{1}{2} r f \partial_{r}\tilde{F}
  \right]
  .
  \label{eq:Moncrief-master-variable-final-sum-l=0}
\end{eqnarray}
This is equivalent to
Eq.~(\ref{eq:Moncrief-master-variable-final-sum-fX=}) with $l=0$ as
\begin{eqnarray}
  f X_{(e)}
  =
  \frac{1-3f}{r} \Phi_{(e)}
  + \frac{1-3f}{4} \tilde{F}
  - \frac{1}{2} r f \partial_{r}\tilde{F}
  .
  \label{eq:Moncrief-master-variable-final-sum-fX=-l=0}
\end{eqnarray}
As in the case of $l\neq 0$ mode, this equation yields the component
$X_{(e)}$ of the metric perturbation is determined by
$(\tilde{F},\Phi_{(e)})$.


The crucial difference between the $l=0$ mode and $l\neq 0$ modes is
Eqs.~(\ref{eq:-Ein-non-vac-XF-constraint-apha-Phie-with-SLambdaF-sum})
and (\ref{eq:ll+1fYe-1st-pert-Ein-non-vac-tr-FAB-div-t-r-sum-3}).
In the $l=0$ case, these equations yield
\begin{eqnarray}
  &&
     \partial_{r}\left[(1-3f) \Phi_{(e)}\right]
     =
     - \frac{8\pi r^{2}}{f} \tilde{T}_{tt}
     ,
     \label{eq:-Ein-non-vac-XF-constraint-Phie-reduce-l=0}
  \\
  &&
     \partial_{t}\left[
     (1-3f) \Phi_{(e)}
     \right]
     =
     - 8 \pi r^{2} f \tilde{T}_{tr}
     ,
     \label{eq:ll+1fYe-reduced-l=0-mode}
\end{eqnarray}
where we used
Eq.~(\ref{eq:Moncrief-master-variable-final-sum-fX=-l=0}) to derive
Eq.~(\ref{eq:ll+1fYe-reduced-l=0-mode}).


The components of the divergence of the energy momentum tensor are
summarized as
\begin{eqnarray}
  &&
     \partial_{t}\tilde{T}_{tt}
     -  f^{2} \partial_{r}\tilde{T}_{rt}
     -  \frac{(1+f)f}{r} \tilde{T}_{rt}
     =
     0
     ,
     \label{eq:div-barTab-linear-AB-t-comp-reduce-l=0}
  \\
  &&
     \partial_{t}\tilde{T}_{tr}
     -  \frac{1-f}{2rf} \tilde{T}_{tt}
     -  f^{2} \partial_{r}\tilde{T}_{rr}
     -  \frac{(3+f)f}{2r} \tilde{T}_{rr}
     =
     0
     ,
     \label{eq:div-barTab-linear-AB-r-comp-reduce-l=0}
  \\
  &&
     \tilde{T}_{(e0)} = 0
     .
     \label{eq:div-barTab-linear-p-mode-reduce-l=0}
\end{eqnarray}
Here, we check the integrability condition of
Eqs.~(\ref{eq:-Ein-non-vac-XF-constraint-Phie-reduce-l=0}) and
(\ref{eq:ll+1fYe-reduced-l=0-mode}).
Differentiating
Eq.~(\ref{eq:-Ein-non-vac-XF-constraint-Phie-reduce-l=0}) with respect
to $t$ and differentiating Eq.~(\ref{eq:ll+1fYe-reduced-l=0-mode}) with
respect to $r$, we obtain the integrability condition of
Eqs.~(\ref{eq:-Ein-non-vac-XF-constraint-Phie-reduce-l=0}) and
(\ref{eq:ll+1fYe-reduced-l=0-mode}) follows
\begin{eqnarray}
  \label{eq:integrability-1-3fPhie}
  0
  &=&
      \partial_{t}\left(- 8 \pi \frac{r^{2}}{f} \tilde{T}_{tt}\right)
      -
      \partial_{r}\left(- 8 \pi r^{2} f \tilde{T}_{tr}\right)
      \nonumber\\
  &=&
      -  8 \pi r^{2} \frac{1}{f} \left[
      \partial_{t}\tilde{T}_{tt}
      -  f^{2} \partial_{r}\tilde{T}_{tr}
      -  \frac{(1+f)f}{r} \tilde{T}_{tr}
      \right]
      .
\end{eqnarray}
This coincides with the component
(\ref{eq:div-barTab-linear-AB-t-comp-reduce-l=0}) of the continuity
equation of the matter field.
Thus, Eqs.~(\ref{eq:-Ein-non-vac-XF-constraint-Phie-reduce-l=0}) and
(\ref{eq:ll+1fYe-reduced-l=0-mode}) are integrable and there exist the
solution $\Phi_{(e)}=\Phi_{(e)}[T_{tt},T_{tr}]$ to these equations.


In the case of $l=0$ mode, the evolution equation
(\ref{eq:Zerilli-Moncrief-eq-final-sum}) has the same form, but the
potential $V_{even}$ defined by
Eq.~(\ref{eq:Zerilli-Moncrief-master-potential-final-sum}) with $l=0$
is given by
\begin{eqnarray}
  V_{even}
  =
  \frac{3(1-f)(1+3f^{2})}{r^{2}(1-3f)^{2}}
  \label{eq:Zerilli-Moncrief-master-potential-reduce-l=0}
\end{eqnarray}
and the source term in Eq.~(\ref{eq:SPhie-def-explicit-sum}) is given
by
\begin{eqnarray}
  S_{(\Phi_{(e)})}
  =
  -  \tilde{T}_{tt}
  -  \frac{r}{2} \partial_{r}\tilde{T}_{tt}
  + \frac{r}{2} \partial_{t}\tilde{T}_{tr}
  - \frac{3(1-f)}{1-3f} \tilde{T}_{tt}
  .
  \label{eq:SPhie-def-explicit-2-sum-l=0}
\end{eqnarray}
Through Eqs.~(\ref{eq:-Ein-non-vac-XF-constraint-Phie-reduce-l=0}) and
(\ref{eq:ll+1fYe-reduced-l=0-mode}), we obtain
\begin{eqnarray}
  &&
  -  \frac{1}{f} \partial_{t}^{2}\Phi_{(e)}
  + \partial_{r}(f \partial_{r}\Phi_{(e)})
  - V_{even} \Phi_{(e)}
     \nonumber\\
  &=&
  \frac{16\pi r}{1-3f}
  \left[
  -  \tilde{T}_{tt}
  -  \frac{r}{2} \partial_{r}\tilde{T}_{tt}
  + \frac{r}{2} \partial_{t}\tilde{T}_{tr}
  - \frac{3(1-f)}{1-3f} \tilde{T}_{tt}
  \right]
  .
  \label{eq:ll+1fYe-reduced-l=0-mode-Ein-non-vac-XF-constraint-Phie-reduce-l=0}
\end{eqnarray}
This coincides with the master equation
(\ref{eq:Zerilli-Moncrief-eq-final-sum}) with $l=0$.
Thus, the master equation (\ref{eq:Zerilli-Moncrief-eq-final-sum})
does not give us any information other than that of
Eqs.~(\ref{eq:-Ein-non-vac-XF-constraint-Phie-reduce-l=0}) and
(\ref{eq:ll+1fYe-reduced-l=0-mode}).


As in the case of $l\neq 0$ modes, the metric component $X_{(e)}$ is
determined by the variables $(\tilde{F},\Phi_{(e)})$ as seen in
Eq.~(\ref{eq:Moncrief-master-variable-final-sum-fX=-l=0}).
Although $\tilde{F}$ is determined by
Eq.~(\ref{eq:-Ein-non-vac-XF-constraint-apha-Phie-with-SLambdaF-sum})
in the $l\neq 0$ case, this is impossible in the $l=0$ case.
Therefore, we have to consider Eq.~(\ref{eq:even-mode-tildeF-eq-Phie})
for the variable $\tilde{F}$ which is trivial in the $l\neq 0$ case
\begin{eqnarray}
  -  \frac{1}{f} \partial_{t}^{2}\tilde{F}
  + \partial_{r}( f \partial_{r}\tilde{F} )
  + \frac{1}{r^{2}} 3(1-f) \tilde{F}
  + \frac{4}{r^{3}} (1-3f) \Phi_{(e)}
  =
  16 \pi \left(
  -  \frac{1}{f} \tilde{T}_{tt}
  + f \tilde{T}_{rr}
  \right)
  .
  \label{eq:even-mode-tildeF-eq-Phie-reduce-l=0}
\end{eqnarray}
This equation has the same form of the inhomogeneous version of the
Regge-Wheeler equation with $l=0$, while the original Regge-Wheeler
equation is valid only for the $l\geq 2$ modes.
If we solve this equation
(\ref{eq:even-mode-tildeF-eq-Phie-reduce-l=0}), we can determine the
variable $\tilde{F}$ which depends on the variable $\Phi_{(e)}$ and
the matter fields $\tilde{T}_{tt}$ and $\tilde{T}_{rr}$.
Then, through this solution
$\tilde{F}=\tilde{F}[\Phi_{(e)},\bar{T}_{tt},\bar{T}_{rr}]$ and the
solution to Eqs.~(\ref{eq:-Ein-non-vac-XF-constraint-Phie-reduce-l=0})
and (\ref{eq:ll+1fYe-reduced-l=0-mode}), we can obtain the variable
$X_{(e)}$ through
Eq.~(\ref{eq:Moncrief-master-variable-final-sum-fX=-l=0}) as a
solution to the linearized Einstein equation for the $l=0$ mode.


The remaining component to be obtained is the component $Y_{(e)}$ of
the metric perturbation.
To obtain the variable $Y_{(e)}$, we remind the original initial value
constraints (\ref{eq:even-FAB-divergence-3-t-comp}) and
(\ref{eq:even-FAB-divergence-3-r-comp}).
In the $l=0$ mode case, the source term $S_{(ec)t}$ and $S_{(ec)r}$
are given by $S_{(ec)t}=S_{(ec)r}=0$ from
Eqs.~(\ref{eq:SecA-def}) and
(\ref{eq:l=0-regularized-energy-momentum-tensor-equivalents}).
Then, the initial value constraints
(\ref{eq:even-FAB-divergence-3-t-comp}) and
(\ref{eq:even-FAB-divergence-3-r-comp}) are given by
\begin{eqnarray}
  &&
     \partial_{r}(fY_{(e)})
     =
     \frac{1}{2} \partial_{t}\tilde{F}
     - \partial_{t}(X_{(e)})
     ,
     \label{eq:even-FAB-divergence-3-t-comp-l=0}
  \\
  &&
     \partial_{t}(fY_{(e)})
     =
     - f \partial_{r}(fX_{(e)})
     -  \frac{1}{2} f^{2} \partial_{r}\tilde{F}
     .
     \label{eq:even-FAB-divergence-3-r-comp-l=0}
\end{eqnarray}
We may regard that Eqs.~(\ref{eq:even-FAB-divergence-3-t-comp-l=0})
and (\ref{eq:even-FAB-divergence-3-r-comp-l=0}) are equations to
obtain the variable $Y_{(e)}$.
Actually, the integrability of these equations is guaranteed by
Eqs.~(\ref{eq:Moncrief-master-variable-final-sum-fX=-l=0}),
(\ref{eq:-Ein-non-vac-XF-constraint-Phie-reduce-l=0}),
(\ref{eq:ll+1fYe-reduced-l=0-mode}),
(\ref{eq:div-barTab-linear-AB-r-comp-reduce-l=0}), and
(\ref{eq:even-mode-tildeF-eq-Phie-reduce-l=0}).
Then, we can obtain the component $Y_{(e)}$ of the metric perturbation
by the direct integration of
Eqs.~(\ref{eq:even-FAB-divergence-3-t-comp-l=0}) and
(\ref{eq:even-FAB-divergence-3-r-comp-l=0}).


We may carry out the above strategy to obtain the $l=0$ mode solution
to the linearized Einstein equations, but it is instructive to
consider the vacuum case where all components of the linearized
energy-momentum tensor ${}^{(1)}\!\ScrT_{ab}$ vanishes before the
derivation of the non-vacuum case.


\subsection{$l=0$ mode vacuum case}
\label{sec:l=0_Schwarzschild_Background-vac}


Here, we consider the vacuum case of the above equations for $l=0$
mode perturbations.
First, we consider
Eqs.~(\ref{eq:-Ein-non-vac-XF-constraint-Phie-reduce-l=0}) and
(\ref{eq:ll+1fYe-reduced-l=0-mode}) with the vacuum condition:
\begin{eqnarray}
  &&
     \partial_{r}\left[(1-3f) \Phi_{(e)}\right] = 0
     ,
     \label{eq:-Ein-non-vac-XF-constraint-Phie-reduce-l=0-vac}
  \\
  &&
     \partial_{t}\left[(1-3f) \Phi_{(e)}\right] = 0
     .
     \label{eq:ll+1fYe-reduced-l=0-vac}
\end{eqnarray}
These equations are easily integrated as
\begin{eqnarray}
  \label{eq:Phie-vac-sol-l=0}
  (1-3f) \Phi_{(e)} = - 2M_{1}, \quad M_{1} \in \RF.
\end{eqnarray}
Furthermore, the variable $\tilde{F}$ is determined by
Eq.~(\ref{eq:even-mode-tildeF-eq-Phie-reduce-l=0}) with vacuum
condition:
\begin{eqnarray}
  -  \frac{1}{f} \partial_{t}^{2}\tilde{F}
  + \partial_{r}( f \partial_{r}\tilde{F} )
  + \frac{1}{r^{2}} 3(1-f) \tilde{F}
  -  \frac{8M_{1}}{r^{3}}
  =
  0
  .
  \label{eq:even-mode-tildeF-eq-Phie-reduce-l=0-vac}
\end{eqnarray}
From Eqs.~(\ref{eq:Moncrief-master-variable-final-sum-fX=-l=0}) and
(\ref{eq:Phie-vac-sol-l=0}), we obtain the component $X_{(e)}$ of the
metric perturbation as follows:
\begin{eqnarray}
  f X_{(e)}
  =
  - \frac{2M_{1}}{r}
  + \frac{1-3f}{4} \tilde{F}
  - \frac{1}{2} r f \partial_{r}\tilde{F}
  .
  \label{eq:Moncrief-master-variable-final-sum-fX=-l=0-vac}
\end{eqnarray}
Moreover, the components $Y_{(e)}$ is obtain the direct integration of
Eqs.~(\ref{eq:even-FAB-divergence-3-t-comp-l=0}) and
(\ref{eq:even-FAB-divergence-3-r-comp-l=0}), because the integrability
is already guaranteed.
Substituting
Eq.~(\ref{eq:Moncrief-master-variable-final-sum-fX=-l=0-vac}) into
Eqs.~(\ref{eq:even-FAB-divergence-3-t-comp-l=0}) and
(\ref{eq:even-FAB-divergence-3-r-comp-l=0}), we obtain
\begin{eqnarray}
  &&
     f \partial_{r}(fY_{(e)})
     =
     - \frac{1}{4} (1-5f) \partial_{t}\tilde{F}
     + \frac{1}{2} r f \partial_{r}\partial_{t}\tilde{F}
     ,
     \label{eq:even-FAB-divergence-3-t-comp-l=0-vac-tmp}
  \\
  &&
     \partial_{t}(fY_{(e)})
     =
     \frac{2M_{1}f}{r^{2}}
     -  \frac{3}{4r} f (1-f) \tilde{F}
     -  \frac{1}{4} f (1-3f) \partial_{r}\tilde{F}
     + \frac{1}{2} r \partial_{t}^{2}\tilde{F}
     ,
     \label{eq:even-FAB-divergence-3-r-comp-l=0-vac-tmp}
\end{eqnarray}
where we used Eq.~(\ref{eq:even-mode-tildeF-eq-Phie-reduce-l=0-vac}).


Here, we assume the existence of the solution to
Eq.~(\ref{eq:even-mode-tildeF-eq-Phie-reduce-l=0-vac}) and we
denote this solution by
\begin{eqnarray}
  \label{eq:tildeF-partialtUpsilon}
  \tilde{F} =: \partial_{t}\Upsilon,
\end{eqnarray}
for our convention.
Substituting Eq.~(\ref{eq:tildeF-partialtUpsilon}) into
Eq.~(\ref{eq:even-mode-tildeF-eq-Phie-reduce-l=0-vac}) and integrating by
$t$, we obtain
\begin{eqnarray}
  -  \frac{1}{f} \partial_{t}^{2}\Upsilon
  + \partial_{r}( f \partial_{r}\Upsilon )
  + \frac{1}{r^{2}} 3(1-f) \Upsilon
  -  \frac{8M_{1}}{r^{3}} t
  + \zeta(r)
  =
  0
  .
  \label{eq:even-mode-tildeF-eq-Phie-reduce-l=0-Upsilon}
\end{eqnarray}
where $\zeta(r)$ is an arbitrary function of $r$.
Using Eq.~(\ref{eq:tildeF-partialtUpsilon}) and integrating by $t$,
Eq.~(\ref{eq:even-FAB-divergence-3-r-comp-l=0-vac-tmp}) yields
\begin{eqnarray}
  fY_{(e)}
  =
  \frac{2M_{1}f}{r^{2}} t
  -  \frac{3}{4r} f (1-f) \Upsilon
  -  \frac{1}{4} f (1-3f) \partial_{r}\Upsilon
  + \frac{1}{2} r \partial_{t}^{2}\Upsilon
  + \Xi(r)
  ,
  \label{eq:even-FAB-divergence-3-r-comp-l=0-vac-tmp-2}
\end{eqnarray}
where $\Xi(r)$ is an arbitrary function of $r$.
Substituting Eq.~(\ref{eq:even-FAB-divergence-3-r-comp-l=0-vac-tmp-2})
into Eq.~(\ref{eq:even-FAB-divergence-3-t-comp-l=0-vac-tmp}) and using
Eq.~(\ref{eq:even-mode-tildeF-eq-Phie-reduce-l=0-Upsilon}), we
obtain
\begin{eqnarray}
  \zeta(r)
  =
  - \frac{4}{1-3f} \partial_{r}\Xi(r)
  .
  \label{eq:partialrXir-zetar-relation}
\end{eqnarray}


In summary, we have obtained the components of $X_{(e)}$, $Y_{(e)}$,
and $\tilde{F}$ of the metric perturbations as follows:
\begin{eqnarray}
  f X_{(e)}
  &=&
  - \frac{2M_{1}}{r}
  + \frac{1-3f}{4} \partial_{t}\Upsilon
  - \frac{1}{2} r f \partial_{r}\partial_{t}\Upsilon
  ,
  \label{eq:Moncrief-master-variable-final-sum-fX=-l=0-vac-sum}
  \\
  fY_{(e)}
  &=&
  \frac{2M_{1}f}{r^{2}} t
  -  \frac{3}{4r} f (1-f) \Upsilon
  -  \frac{1}{4} f (1-3f) \partial_{r}\Upsilon
  + \frac{1}{2} r \partial_{t}^{2}\Upsilon
  + \Xi(r)
  ,
  \label{eq:even-FAB-divergence-3-r-comp-l=0-vac-tmp-2-sum}
\end{eqnarray}
and
\begin{eqnarray}
  \label{eq:tildeF-partialtUpsilon-sum}
  \tilde{F} = \partial_{t}\Upsilon, \quad
  -  \frac{1}{f} \partial_{t}^{2}\Upsilon
  + \partial_{r}( f \partial_{r}\Upsilon )
  + \frac{1}{r^{2}} 3(1-f) \Upsilon
  -  \frac{8M_{1}}{r^{3}} t
  - \frac{4}{1-3f} \partial_{r}\Xi(r)
  =
  0
  ,
\end{eqnarray}
and $\Xi(r)$ is an arbitrary function of $r$.


Here, we consider the covariant form $\ScrF_{ab}$ of the $l=0$ mode
metric perturbation.
According to
Proposal~\ref{proposal:treatment-proposal-on-pert-on-spherical-BG}, we
impose the regularity on $S^{2}$ to the harmonic function
$k_{(\Delta)}$ so that
\begin{eqnarray}
  \label{eq:regularized-kDelta}
  k_{(\Delta)} = 1.
\end{eqnarray}
Since $\tilde{F}_{D}^{\;\;\;D}=0$ by
Eq.~(\ref{eq:eventildeFtrace-matter-reduce-2-l=0}) for $l=0$ mode
perturbations, the gauge-invariant metric perturbation $\ScrF_{ab}$
for the $l=0$ mode is given by
\begin{eqnarray}
  \ScrF_{ab}
  &=&
      \tilde{F}_{AB} (dx^{A})_{a}(dx^{B})_{b}
      +
      \frac{1}{2} \gamma_{pq} r^{2} \tilde{F} (dx^{p})_{a} (dx^{q})_{b}
      \nonumber\\
  &=&
      - (f X_{(e)}) \left\{ (dt)_{a}(dt)_{b} + f^{-2} (dr)_{a}(dr)_{b}\right\}
      + 2 (f Y_{(e)})  f^{-1} (dt)_{(A}(dr)_{B)}
      \nonumber\\
  &&
      + \frac{1}{2} \gamma_{pq} r^{2} \tilde{F} (dx^{p})_{a} (dx^{q})_{b}
      .
      \label{eq:2+2-gauge-inv-var-even-cov-mode-l=0-sum}
\end{eqnarray}


As in the case of the $l=1$ odd-mode perturbation in Part
I paper~\cite{K.Nakamura-2021c}, the solutions
(\ref{eq:Moncrief-master-variable-final-sum-fX=-l=0-vac-sum})--(\ref{eq:tildeF-partialtUpsilon-sum})
may include the terms in the form of ${\pounds}_{V}g_{ab}$ for a
vector field $V^{a}$.
To find the term ${\pounds}_{V}g_{ab}$, we consider the generator
$V_{a}$ whose components are given by
\begin{eqnarray}
  V_{a} = V_{t}(t,r)(dt)_{a} + V_{r}(r,t)(dr)_{a}.
\end{eqnarray}
Then, the nonvanishing components of  ${\pounds}_{V}g_{ab}$ are given
by
\begin{eqnarray}
  \label{eq:poundsVgab-components-tt-even-l=0}
  &&
     {\pounds}_{V}g_{tt}
     =
     2 \partial_{t}V_{t} - f f' V_{r}
     ,
  \\
  \label{eq:poundsVgab-components-tr-even-l=0}
  &&
     {\pounds}_{V}g_{tr}
     =
     \partial_{t}V_{r} + \partial_{r}V_{t} - \frac{f'}{f} V_{t}
     ,
  \\
  \label{eq:poundsVgab-components-rr-even-l=0}
  &&
     {\pounds}_{V}g_{rr}
     =
     2 \partial_{r}V_{r} + \frac{f'}{f} V_{r}
     ,
  \\
  \label{eq:poundsVgab-components-thetatheta-even-l=0}
  &&
     {\pounds}_{V}g_{\theta\theta}
     =
     2 rf V_{r}
     ,
  \\
  \label{eq:poundsVgab-components-phiphi-even-l=0}
  &&
     {\pounds}_{V}g_{\phi\phi}
     =
     2 rf \sin^{2}\theta V_{r}
     .
\end{eqnarray}
From Eqs.~(\ref{eq:2+2-gauge-inv-var-even-cov-mode-l=0-sum}),
(\ref{eq:poundsVgab-components-thetatheta-even-l=0}), and
(\ref{eq:poundsVgab-components-phiphi-even-l=0}), we choose
\begin{eqnarray}
  \label{eq:Vr-tildeF-original-relation}
  V_{r} = \frac{1}{4f} r \tilde{F} = \frac{1}{4f} r \partial_{t}\Upsilon, \quad
  {\pounds}_{V}g_{\theta\theta}
  = \frac{1}{\sin^{2}\theta} {\pounds}_{V}g_{\phi\phi}
  = \frac{1}{2} r^{2} \tilde{F}= \frac{1}{2} r^{2} \partial_{t}\Upsilon.
\end{eqnarray}
Substituting Eq.~(\ref{eq:Vr-tildeF-original-relation}) into
Eqs.~(\ref{eq:poundsVgab-components-tt-even-l=0})--(\ref{eq:poundsVgab-components-rr-even-l=0}),
we obtain
\begin{eqnarray}
  \label{eq:poundsVgab-components-tt-even-l=0-red}
  &&
     {\pounds}_{V}g_{tt}
     =
     2 \partial_{t}V_{t}
     -
     \frac{1}{4} (1-f) \partial_{t}\Upsilon
     ,
  \\
  \label{eq:poundsVgab-components-tr-even-l=0-red}
  &&
     {\pounds}_{V}g_{tr}
     =
     \frac{1}{4f} r \partial_{t}^{2}\Upsilon
     +
     \partial_{r}V_{t}
     -
     \frac{1}{fr} (1-f) V_{t}
     ,
  \\
  \label{eq:poundsVgab-components-ttheta-even-l=0-red}
  &&
     {\pounds}_{V}g_{rr}
     =
     -  \frac{1}{4f^{2}} (1-3f) \partial_{t}\Upsilon
     + \frac{1}{2f} r \partial_{r}\partial_{t}\Upsilon
     .
\end{eqnarray}
To identify the degree of freedom which expressed as
${\pounds}_{V}g_{ab}$ in $X_{(e)}$, we choose
\begin{eqnarray}
  \label{eq:Vt-l=0-vac-choice}
  \partial_{t}V_{t}
  =
  \frac{1}{4} f \partial_{t}\Upsilon
  + \frac{1}{4} r f \partial_{r}\partial_{t}\Upsilon
\end{eqnarray}
so that
\begin{eqnarray}
  \label{eq:poundsVgab-tt-even-l=0-Vt-used}
  {\pounds}_{V}g_{tt}
  =
  -  \frac{1}{4} (1-3f) \partial_{t}\Upsilon
  + \frac{1}{2} r f \partial_{r}\partial_{t}\Upsilon
  .
\end{eqnarray}
Then, we obtain
\begin{eqnarray}
  V_{t}
  =
  \frac{1}{4} f \Upsilon
  + \frac{1}{4} r f \partial_{r}\Upsilon
  + \gamma(r)
  ,
  \label{eq:Vt-generator-component-l=0}
\end{eqnarray}
where $\gamma(r)$ is an arbitrary function of $r$.
Substituting Eq.~(\ref{eq:Vt-generator-component-l=0}) into
Eq.~(\ref{eq:poundsVgab-components-tr-even-l=0-red}) and using
the equation (\ref{eq:tildeF-partialtUpsilon-sum}) for $\Upsilon$, we
obtain
\begin{eqnarray}
  {\pounds}_{V}g_{tr}
  &=&
      \frac{2M_{1}}{r^{2}} t
      + \frac{r}{2f} \partial_{t}^{2}\Upsilon
      -  \frac{1}{4} (1-3f) \partial_{r}\Upsilon
      -  \frac{3}{4r} (1-f) \Upsilon
      \nonumber\\
  &&
     + \frac{r}{1-3f} \partial_{r}\Xi(r)
     + \partial_{r}\gamma(r)
     -  \frac{1}{fr} (1-f) \gamma(r)
     .
     \label{eq:poundsVgab-tr-even-l=0-Vt-tildeF-used}
\end{eqnarray}
From the solutions
(\ref{eq:Moncrief-master-variable-final-sum-fX=-l=0-vac-sum}),
(\ref{eq:even-FAB-divergence-3-r-comp-l=0-vac-tmp-2-sum}), and
(\ref{eq:tildeF-partialtUpsilon-sum}), and the expression
(\ref{eq:2+2-gauge-inv-var-even-cov-mode-l=0-sum}) of the
gauge-invariant part of the metric perturbation, and the components
(\ref{eq:Vr-tildeF-original-relation}),
(\ref{eq:poundsVgab-components-ttheta-even-l=0-red}),
(\ref{eq:poundsVgab-tt-even-l=0-Vt-used}), and
(\ref{eq:poundsVgab-tr-even-l=0-Vt-tildeF-used}) of
${\pounds}_{V}g_{ab}$, we obtain
\begin{eqnarray}
  \ScrF_{ab}
  &=&
      \frac{2M_{1}}{r}
      \left(
      (dt)_{a}(dt)_{b}
      +
      f^{-2}(dr)_{a}(dr)_{b}
      \right)
      + {\pounds}_{V}g_{ab}
      \nonumber\\
  &&
      + 2 \left(
      \frac{1}{f} \Xi(r)
      -  \frac{r}{1-3f} \partial_{r}\Xi(r)
      -  \partial_{r}\gamma(r)
      +  \frac{1}{fr} (1-f) \gamma(r)
      \right)  (dt)_{(a}(dr)_{b)}
      .
      \label{eq:2+2-gauge-inv-var-even-cov-mode-l=0-sum-tmp}
\end{eqnarray}
As a choice of the generator $V_{a}$, we choose the arbitrary function
$\gamma(r)$ in $V_{a}$ such that
\begin{eqnarray}
  \gamma(r)
  =
  -  \frac{r}{(1-3f)} \Xi(r)
  + f \int dr \frac{2}{f(1-3f)^{2}} \Xi(r)
  .
  \label{eq:l=0-generator-gamma-choice}
\end{eqnarray}
Then, we obtain
\begin{eqnarray}
  \ScrF_{ab}
  &=&
      \frac{2M_{1}}{r}
      \left(
      (dt)_{a}(dt)_{b}
      +
      f^{-2}(dr)_{a}(dr)_{b}
      \right)
      + {\pounds}_{V}g_{ab}
      ,
      \label{eq:2+2-gauge-inv-var-even-cov-mode-l=0-vac-final}
\end{eqnarray}
where
\begin{eqnarray}
  \label{eq:l=0-vac-generator-sum}
  V_{a}
  =
  \left(
  \frac{f}{4} \Upsilon
  + \frac{rf}{4} \partial_{r}\Upsilon
  -  \frac{r}{1-3f} \Xi(r)
  + f \int dr \frac{2}{f(1-3f)^{2}} \Xi(r)
  \right)
  (dt)_{a}
  +
  \frac{r}{4f} \partial_{t}\Upsilon
  (dr)_{a}
  .
\end{eqnarray}
The function $\Upsilon(t,r)$ is the solution to the second equation
(\ref{eq:tildeF-partialtUpsilon-sum}).


The solution (\ref{eq:2+2-gauge-inv-var-even-cov-mode-l=0-vac-final})
is the $O(\epsilon)$ mass parameter perturbation $M+\epsilon M_{1}$
of the Schwarzschild spacetime apart from the term the Lie derivative
of the background metric $g_{ab}$.
Since $l=0$ mode is the spherically symmetric perturbations, the
solution (\ref{eq:2+2-gauge-inv-var-even-cov-mode-l=0-vac-final}) is
the realization of the linearized gauge-invariant version of
Birkhoff's theorem~\cite{S.W.Hawking-G.F.R.Ellis-1973}.
We also note that the vector field $V_{a}$ is also gauge-invariant in
the sense of the second kind.
Here, we have to emphasize that the generator
(\ref{eq:l=0-vac-generator-sum}) with the second equation in
Eq.~(\ref{eq:tildeF-partialtUpsilon-sum}) is necessary if we include
the perturbative Schwarzschild mass parameter $M_{1}$ as the solution
to the linearized Einstein equation in our framework.
This can be seen from the second equation in
Eq.~(\ref{eq:tildeF-partialtUpsilon-sum}).
This equation indicates that $M_{1}=0$ if we choose $\Upsilon =0$ for
arbitrary time $t$.


\subsection{$l=0$ mode non-vacuum case}
\label{sec:l=0_Schwarzschild_Background-non-vac-main}


Inspecting the above vacuum case, we apply the method of variational
constants.
In Eq.~(\ref{eq:Phie-vac-sol-l=0}), the Schwarzschild mass parameter
perturbation $M_{1}$ is an integration constant.
Then, we choose the function $m_{1}(t,r)$ so that
\begin{eqnarray}
  \label{eq:l=0-mass1function-def}
     m_{1}(t,r) := - \frac{1}{2} (1-3f) \Phi_{(e)}.
\end{eqnarray}
The integrability of Eqs.~(\ref{eq:-Ein-non-vac-XF-constraint-Phie-reduce-l=0}) and
(\ref{eq:ll+1fYe-reduced-l=0-mode}) was already confirmed in
Eq.~(\ref{eq:integrability-1-3fPhie}).
Then, we obtain
\begin{eqnarray}
  m_{1}(t,r)
  =
  4 \pi \int dr \frac{r^{2}}{f} \tilde{T}_{tt}
  +
  M_{1}
  =
  4 \pi \int dt r^{2} f \tilde{T}_{tr}
  +
  M_{1}
  .
  \label{eq:l=0-mode-m1-sol}
\end{eqnarray}
Equation~(\ref{eq:Moncrief-master-variable-final-sum-fX=-l=0}) yields
the component $X_{(e)}$ of the metric perturbation as follows:
\begin{eqnarray}
  f X_{(e)}
  =
  - \frac{2m_{1}(t,r)}{r}
  + \frac{1-3f}{4} \tilde{F}
  - \frac{1}{2} r f \partial_{r}\tilde{F}
  .
  \label{eq:Moncrief-master-variable-final-sum-fX=-l=0-m1}
\end{eqnarray}
As discussed in above, the variable $\tilde{F}$ is determined by
Eq.~(\ref{eq:even-mode-tildeF-eq-Phie-reduce-l=0}).
As in the vacuum case in
Sec.~\ref{sec:l=0_Schwarzschild_Background-vac}, we introduce the
function $\Upsilon$ such that
\begin{eqnarray}
  &&
     \tilde{F} =: \partial_{t}\Upsilon,
     \label{eq:tildeF-def-partialt-Upsilon-non-vac}
  \\
  &&
     -  \frac{1}{f} \partial_{t}^{2}\Upsilon
     + \partial_{r}( f \partial_{r}\Upsilon )
     + \frac{3(1-f)}{r^{2}} \Upsilon
     - \frac{8}{r^{3}} \int dt m_{1}(t,r)
     + \zeta(r)
     \nonumber\\
     &=&
     16 \pi \int dt \left(
     - \frac{1}{f} \tilde{T}_{tt}
     + f \tilde{T}_{rr}
     \right)
     ,
  \label{eq:even-mode-tildeF-eq-Phie-reduce-l=0-m1}
\end{eqnarray}
where $\zeta(r)$ is an arbitrary function of $r$.
Through the variable $\Upsilon$ and
Eq.~(\ref{eq:Moncrief-master-variable-final-sum-fX=-l=0-m1}),
Eq.~(\ref{eq:even-FAB-divergence-3-r-comp-l=0}) is integrated as
follows:
\begin{eqnarray}
  f Y_{(e)}
  &=&
      \frac{2f}{r^{2}} \int dt m_{1}(t,r)
      + 8 \pi r f^{2} \int dt \tilde{T}_{rr}
      \nonumber\\
  &&
     -  \frac{3f(1-f)}{4r} \Upsilon
     -  \frac{f(1-3f)}{4} \partial_{r}\Upsilon
     + \frac{r}{2} \partial_{t}^{2}\Upsilon
     + \Xi(r)
     ,
  \label{eq:even-FAB-divergence-3-r-comp-l=0-int-m1}
\end{eqnarray}
where $\Xi(r)$ is an different arbitrary function of $r$ from
$\zeta(r)$.
Substituting Eq.~(\ref{eq:even-FAB-divergence-3-r-comp-l=0-int-m1})
into Eq.~(\ref{eq:even-FAB-divergence-3-t-comp-l=0}), and using
Eqs.~(\ref{eq:l=0-mode-m1-sol}),
(\ref{eq:Moncrief-master-variable-final-sum-fX=-l=0-m1}),
(\ref{eq:even-mode-tildeF-eq-Phie-reduce-l=0-m1}), and the component
(\ref{eq:div-barTab-linear-AB-r-comp-reduce-l=0}) of the continuity
equation, we obtain
\begin{eqnarray}
  \zeta(r)
  &=&
      - \frac{4}{1-3f} \partial_{r}\Xi(r)
      \label{eq:partialrXir-zetar-relation-non-vac}
\end{eqnarray}
as expected from the vacuum case in
Sec.~\ref{sec:l=0_Schwarzschild_Background-vac}.


In summary, we have obtained the solution to the components of the
metric perturbations $X_{(e)}$, $Y_{(e)}$, and $\tilde{F}$ as follows:
\begin{eqnarray}
     f X_{(e)}
     &=&
     - \frac{2m_{1}(t,r)}{r}
     + \frac{1-3f}{4} \partial_{t}\Upsilon
     - \frac{1}{2} r f \partial_{r}\partial_{t}\Upsilon
     ,
     \label{eq:Moncrief-master-variable-final-sum-fX=-l=0-m1-sum}
  \\
     f Y_{(e)}
     &=&
     \frac{2f}{r^{2}} \int dt m_{1}(t,r)
     + 8 \pi r f^{2} \int dt \tilde{T}_{rr}
     \nonumber\\
  &&
     -  \frac{3f(1-f)}{4r} \Upsilon
     -  \frac{f(1-3f)}{4} \partial_{r}\Upsilon
     + \frac{r}{2} \partial_{t}^{2}\Upsilon
     + \Xi(r)
     ,
     \label{eq:even-FAB-divergence-3-r-comp-l=0-int-m1-sum}
  \\
     \tilde{F} &=:& \partial_{t}\Upsilon
     ,
     \label{eq:tildeF-def-partialt-Upsilon-non-vac-sum}
  \\
  &&
     \partial_{t}^{2}\Upsilon
     - f \partial_{r}( f \partial_{r}\Upsilon )
     - \frac{3f(1-f)}{r^{2}} \Upsilon
     + \frac{8f}{r^{3}} \int dt m_{1}(t,r)
     + \frac{4f \partial_{r}\Xi(r)}{1-3f}
     \nonumber\\
     &=&
     16 \pi \int dt \left(
      \tilde{T}_{tt}
     - f^{2} \tilde{T}_{rr}
     \right)
     .
     \label{eq:even-mode-tildeF-eq-Phie-reduce-l=0-m1-sum}
\end{eqnarray}


Here, we consider the covariant form of the above $l=0$ mode
non-vacuum solutions.
As in the vacuum case in
Sec.~\ref{sec:l=0_Schwarzschild_Background-vac}, we show the
expression (\ref{eq:2+2-gauge-inv-var-even-cov-mode-l=0-sum}) of the
above non-vacuum solution
\begin{eqnarray}
  \ScrF_{ab}
  &=&
      - ( f X_{(e)} ) \left[(dt)_{a}(dt)_{b} + f^{-2} (dr)_{a}(dr)_{b}\right]
      + 2 ( f Y_{(e)} )  f^{-1} (dt)_{(A}(dr)_{B)}
      \nonumber\\
  &&
      + \frac{1}{2} \gamma_{pq} r^{2} \partial_{t}\Upsilon (dx^{p})_{a} (dx^{q})_{b}
      .
      \label{eq:2+2-gauge-inv-var-even-cov-mode-l=0-non-vac-bare}
\end{eqnarray}
The components of $\ScrF_{ab}$ are given by
\begin{eqnarray}
     \ScrF_{tt}
     &=&
     \frac{2m_{1}(t,r)}{r}
     - \frac{1-3f}{4} \partial_{t}\Upsilon
     + \frac{1}{2} r f \partial_{r}\partial_{t}\Upsilon
     ,
     \label{eq:calFtt-non-vac-sol-l=0}
  \\
     \ScrF_{tr}
     &=&
     \frac{2}{r^{2}} \int dt m_{1}(t,r)
     + 8 \pi r f \int dt \tilde{T}_{rr}
     \nonumber\\
  &&
     -  \frac{3(1-f)}{4r} \Upsilon
     -  \frac{(1-3f)}{4} \partial_{r}\Upsilon
     + \frac{r}{2f} \partial_{t}^{2}\Upsilon
     + \frac{1}{f} \Xi(r)
     ,
     \label{eq:calFtr-non-vac-sol-l=0}
  \\
     \ScrF_{rr}
     &=&
     \frac{2m_{1}(t,r)}{rf^{2}}
     -  \frac{1-3f}{4f^{2}} \partial_{t}\Upsilon
     + \frac{r}{2f} \partial_{r}\partial_{t}\Upsilon
     ,
     \label{eq:calFrr-non-vac-sol-l=0}
  \\
     \ScrF_{\theta\theta}
     &=&
     \frac{r^{2}}{2} \partial_{t}\Upsilon
     =
     \frac{1}{\sin^{2}\theta} \ScrF_{\phi\phi}
     .
     \label{eq:calFthetatheta-calFphiphinon-vac-sol-l=0}
\end{eqnarray}


As in the vacuum case, we consider the term in the form
${\pounds}_{V}g_{ab}$ with the generator
\begin{eqnarray}
  V_{a} = V_{t}(t,r)(dt)_{a} + V_{r}(r,t)(dr)_{a}.
\end{eqnarray}
Then, we obtain
Eqs.~(\ref{eq:poundsVgab-components-tt-even-l=0})--(\ref{eq:poundsVgab-components-phiphi-even-l=0}).
Comparing Eqs.~(\ref{eq:poundsVgab-components-thetatheta-even-l=0}),
(\ref{eq:poundsVgab-components-phiphi-even-l=0}), and
(\ref{eq:calFthetatheta-calFphiphinon-vac-sol-l=0}), we choose $V_{r}$
so that
\begin{eqnarray}
  \label{eq:Vr-choice-non-vac}
  V_{r} = \frac{1}{4f} r \partial_{t}\Upsilon, \quad
  {\pounds}_{V}g_{\theta\theta}
  =
  \frac{1}{\sin^{2}\theta}
  {\pounds}_{V}g_{\phi\phi}
  =
  \frac{1}{2} r^{2} \partial_{t}\Upsilon
  ,
\end{eqnarray}
and we have
\begin{eqnarray}
  \label{eq:calFthetatheta-phiphi-poundsVg-exp}
  \ScrF_{\theta\theta}
  =
  {\pounds}_{V}g_{\theta\theta}
  , \quad
  \ScrF_{\phi\phi}
  =
  {\pounds}_{V}g_{\phi\phi}
  .
\end{eqnarray}
Substituting the choice (\ref{eq:Vr-choice-non-vac}) into
Eq.~(\ref{eq:poundsVgab-components-rr-even-l=0}) and compare with
Eq.~(\ref{eq:calFrr-non-vac-sol-l=0}), we obtain
\begin{eqnarray}
  \label{eq:calFrr-poundsVg-exp}
  {\pounds}_{V}g_{rr}
  =
  -  \frac{1-3f}{4f^{2}} \partial_{t}\Upsilon
  + \frac{1}{2f} r \partial_{r}\partial_{t}\Upsilon
  ,
  \quad
  \ScrF_{rr}
  =
  \frac{2m_{1}(t,r)}{rf^{2}}
  + {\pounds}_{V}g_{rr}
  .
\end{eqnarray}
Substituting the choice $V_{r}$ in Eq.~(\ref{eq:Vr-choice-non-vac})
into Eq.~(\ref{eq:poundsVgab-components-tt-even-l=0}) and comparing
with Eq.~(\ref{eq:calFtt-non-vac-sol-l=0}), we choose
\begin{eqnarray}
  V_{t}
  =
  \frac{1}{4} f \Upsilon
  + \frac{1}{4} r f \partial_{r}\Upsilon
  + \gamma(r)
  ,
  \label{eq:Vt-choice-non-vac-l=0}
\end{eqnarray}
and obtain
\begin{eqnarray}
  \label{eq:calFtt-poundsVg-exp}
  {\pounds}_{V}g_{tt}
  =
  - \frac{1-3f}{4} \partial_{t}\Upsilon
  + \frac{1}{2} r f \partial_{r}\partial_{t}\Upsilon
  , \quad
  \ScrF_{tt}
  =
  \frac{2m_{1}(t,r)}{r}
  + {\pounds}_{V}g_{tt}
  .
\end{eqnarray}


Finally, from Eq.~(\ref{eq:poundsVgab-components-tr-even-l=0}) with
the choice (\ref{eq:Vt-choice-non-vac-l=0}) of $V_{t}$ and the choice
(\ref{eq:Vr-choice-non-vac}) of $V_{r}$, we obtain
\begin{eqnarray}
  {\pounds}_{V}g_{tr}
  &=&
      \frac{1}{4f} r \partial_{t}^{2}\Upsilon
      -  \frac{1-3f}{4} \partial_{r}\Upsilon
      + \frac{1}{4} r \partial_{r}( f \partial_{r}\Upsilon )
      + \partial_{r}\gamma(r)
      - \frac{1-f}{fr} \gamma(r)
      .
\end{eqnarray}
Furthermore, using
Eq.~(\ref{eq:even-mode-tildeF-eq-Phie-reduce-l=0-m1-sum}), we have
\begin{eqnarray}
  {\pounds}_{V}g_{tr}
  &=&
      \frac{2}{r^{2}} \int dt m_{1}(t,r)
      - 4 \pi \frac{r}{f} \int dt \tilde{T}_{tt}
      + 4 \pi r f \int dt \tilde{T}_{rr}
      \nonumber\\
  &&
      + \frac{1}{2f} r \partial_{t}^{2}\Upsilon
      -  \frac{1-3f}{4} \partial_{r}\Upsilon
      -  \frac{3(1-f)}{4r} \Upsilon
      \nonumber\\
  &&
      + \partial_{r}\gamma(r)
      -  \frac{1-f}{fr} \gamma(r)
      + \frac{r \partial_{r}\Xi(r)}{1-3f}
      .
\end{eqnarray}
Through Eq.~(\ref{eq:calFtr-non-vac-sol-l=0}), we obtain
\begin{eqnarray}
  \label{eq:calFtr-poundsVg-exp-tmp}
  \ScrF_{tr}
  &=&
      4 \pi r \int dt \left(
      \frac{1}{f} \tilde{T}_{tt}
      + f \tilde{T}_{rr}
      \right)
      + {\pounds}_{V}g_{tr}
      \nonumber\\
  &&
      + f \left(
      \frac{2}{f(1-3f)^{2}} \Xi(r)
      -  \partial_{r}\left(\frac{r}{f(1-3f)} \Xi(r)\right)
      - \partial_{r}( \frac{1}{f} \gamma(r) )
      \right)
      .
\end{eqnarray}
The same choice of $\gamma(r)$ in the generator $V_{a}$ as
Eq.~(\ref{eq:l=0-generator-gamma-choice}) yields
\begin{eqnarray}
  \label{eq:calFtr-poundsVg-exp}
  \ScrF_{tr}
  &=&
      4 \pi r \int dt \left(
      \frac{1}{f} \tilde{T}_{tt}
      + f \tilde{T}_{rr}
      \right)
      + {\pounds}_{V}g_{tr}
      .
\end{eqnarray}
Thus, we have obtained
\begin{eqnarray}
  \label{eq:calFab+poundsVg-l=0-non-vac-final}
  \ScrF_{ab}
  &=&
      \frac{2}{r} \left(M_{1}+4\pi \int dr \frac{r^{2}}{f} T_{tt}\right)
      \left((dt)_{a}(dt)_{a}+ \frac{1}{f^{2}} (dr)_{a}(dr)_{a}\right)
      \nonumber\\
  &&
      + 2 \left[4 \pi r \int dt \left(\frac{1}{f} \tilde{T}_{tt} + f \tilde{T}_{rr} \right)\right] (dt)_{(a}(dr)_{b)}
      + {\pounds}_{V}g_{ab}
      ,
\end{eqnarray}
where
\begin{eqnarray}
  \label{eq:Va-result-non-vac-final}
  V_{a}
  =
  \left(
  \frac{f}{4} \Upsilon
  + \frac{rf}{4} \partial_{r}\Upsilon
  -  \frac{r \Xi(r)}{(1-3f)}
  + f \int dr \frac{2 \Xi(r)}{f(1-3f)^{2}}
  \right) (dt)_{a}
  +
  \frac{1}{4f} r \partial_{t}\Upsilon (dr)_{a}
  .
\end{eqnarray}
The variable $\Upsilon$ must satisfy
Eq.~(\ref{eq:even-mode-tildeF-eq-Phie-reduce-l=0-m1-sum}).
We also note that the expression of $\ScrF_{ab}$ is not unique,
since we may choose different vector field $V_{a}$.
We can also choose the time component $V_{t}$ of the vector field
$V_{a}$ so that $\ScrF_{tr}={\pounds}_{V}g_{tr}$.
In this case, the additional terms appear in the component
$\ScrF_{tt}$.


We also note that the term ${\pounds}_{V}g_{ab}$ in
Eq.~(\ref{eq:calFab+poundsVg-l=0-non-vac-final}) is gauge-invariant of
the second kind.
Furthermore, unlike the vacuum case, the variable $\Upsilon$ in this
term includes information of the matter field through
Eq.~(\ref{eq:even-mode-tildeF-eq-Phie-reduce-l=0-m1-sum}).
In this sense, the term ${\pounds}_{V}g_{ab}$ in
Eq.~(\ref{eq:calFab+poundsVg-l=0-non-vac-final}) is physical.


\section{$l=1$ mode non-vacuum perturbations on the Schwarzschild Background}
\label{sec:l=1_Schwarzschild_Background-non-vac}


In this section, we consider the $l=1$ mode perturbations based
through the variables defined in
Secs.~\ref{sec:review-of-general-framework-GI-perturbation-theroy} and~\ref{sec:Schwarzschild_Background-non-vaccum-even-treatment}.
Even in the case of $l=1$ mode, the gauge-invariant variables given by
Eqs.~(\ref{eq:2+2-gauge-invariant-variables-calFAB})--(\ref{eq:2+2-gauge-invariant-variables-calFpq})
are valid.
Since the mode-by-mode analyses are possible in our formulation, we can
consider $l=1$ modes, separately.
For the $l=1$ even-mode perturbations, the component $\ScrF_{Ap}$ of the
gauge-invariant part of the metric perturbation vanishes and the other
components are given by
\begin{eqnarray}
  \label{eq:FAB-FAp-F-def-reduce-even-l=1}
  \ScrF_{AB} := \sum_{m=-1}^{1} \tilde{F}_{AB} k_{(\hat{\Delta+2})m}, \quad
  \ScrF_{pq} := \frac{1}{2} \gamma_{pq} r^{2} \sum_{m=-1}^{1} \tilde{F} k_{(\hat{\Delta}+2)m}.
\end{eqnarray}
We can also separate the trace part $\tilde{F}_{D}^{\;\;\;D}$ and the
traceless part $\tilde{\FF}_{AB}$ for the metric perturbation
$\tilde{F}_{AB}$ as Eq.~(\ref{eq:FF-def}).
We also consider the components of the traceless part $\FF_{AB}$ as
Eq.~(\ref{eq:Xe-Ye-def}).


Following
Proposal~\ref{proposal:treatment-proposal-on-pert-on-spherical-BG}, we
impose the regularity to the harmonic function
$k_{(\hat{\Delta}+2)m}$.
Then, we have
\begin{eqnarray}
  \label{eq:regularized-kDelta+2-vanishing-components}
  \left(
  \hat{D}_{p}\hat{D}_{q}
  -
  \frac{1}{2} \gamma_{pq} \hat{\Delta}
  \right)k_{(\hat{\Delta}+2)m}
  =
  \epsilon_{r(p}\hat{D}_{q)}\hat{D}^{r}k_{(\hat{\Delta}+2)m}
  =
  0
  .
\end{eqnarray}
In this case, the only remaining components of the linearized
energy-momentum tensor ${}^{(1)}\!\ScrT_{ab}$ are given by
\begin{eqnarray}
  {}^{(1)}\!\ScrT_{ab}
  &=&
      \sum_{m=-1}^{1} \tilde{T}_{AB} k_{(\hat{\Delta+2})} (dx^{A})_{a} (dx^{B})_{b}
      \nonumber\\
  &&
     +
     2r
     \sum_{m=-1}^{1} \left\{
     \tilde{T}_{(e1)A} \hat{D}_{p}k_{(\hat{\Delta}+2)m}
     +
     \tilde{T}_{(o1)A} \epsilon_{pr}\hat{D}^{r}k_{(\hat{\Delta}+2)m}
     \right\} (dx^{A})_{(a} (dx^{p})_{b)}
     \nonumber\\
  &&
     +
     \frac{1}{2} r^{2} \gamma_{pq} \sum_{m=-1}^{1} \tilde{T}_{(e0)} k_{(\hat{\Delta}+2)m} (dx^{p})_{a} (dx^{q})_{b}
     .
     \label{eq:l=1-regularized-energy-momentum-tensor}
\end{eqnarray}
Therefore, for even-mode perturbations, we can safely regard that
\begin{eqnarray}
  \label{eq:l=1-regularized-energy-momentum-tensor-equivalents}
  \tilde{T}_{(e2)}=0.
\end{eqnarray}
From Eqs.~(\ref{eq:linearized-Einstein-pq-traceless-even}) and
(\ref{eq:l=1-regularized-energy-momentum-tensor-equivalents}), the
components $\tilde{F}_{AB}$ is traceless.
Then, we may concentrate on the components $X_{(e)}$ and $Y_{(e)}$
defined by Eq.~(\ref{eq:Xe-Ye-def}) and the component $\tilde{F}$ as
the metric perturbations.
Furthermore, all arguments in
Sec.~\ref{sec:Schwarzschild_Background-non-vaccum-even-treatment} are
valid even in the case of $l=1$ modes.
Therefore, we may use
Eqs.~(\ref{eq:Moncrief-master-variable-final-sum})--(\ref{eq:even-mode-tildeF-eq-Phie-remainig-source-tmp})
when we derive the $l=1$ mode solutions to the linearized Einstein
equations.


From the definition (\ref{eq:mu-Lambda-defs}) of $\Lambda$, we obtain
\begin{eqnarray}
  \label{eq:mu-Lambda-defs-l=1}
  \Lambda = 3(1-f).
\end{eqnarray}
Then, the Moncrief variable $\Phi_{(e)}$ defined by
Eq.~(\ref{eq:Moncrief-master-variable-final-sum}) is given by
\begin{eqnarray}
  \label{eq:Moncrief-master-variable-final-l=1}
  \Phi_{(e)}
  :=
  \frac{r}{3(1-f)} \left[
  f X_{(e)}
  - \frac{3(1-f)}{4} \tilde{F}
  + \frac{1}{2} r f \partial_{r}\tilde{F}
  \right]
  .
\end{eqnarray}
In other words, the components $X_{(e)}$ is given by
\begin{eqnarray}
  f X_{(e)}
  =
  \frac{3(1-f)}{r} \Phi_{(e)}
  + \frac{3(1-f)}{4} \tilde{F}
  - \frac{1}{2} r f \partial_{r}\tilde{F}
  \label{eq:Moncrief-master-variable-final-sum-fX=l=1}
\end{eqnarray}
as a solution to the linearized Einstein equation, if the variables
$\Phi_{(e)}$ and $\tilde{F}$ are given as solutions to the linearized
Einstein equation.
Furthermore, from
Eqs.~(\ref{eq:-Ein-non-vac-XF-constraint-apha-Phie-with-SLambdaF-sum})
and (\ref{eq:ll+1fYe-1st-pert-Ein-non-vac-tr-FAB-div-t-r-sum-3}), we
obtain
\begin{eqnarray}
  \tilde{F}
  &=&
      -  4 f \partial_{r}\Phi_{(e)}
      -  \frac{4(1-f)}{r} \Phi_{(e)}
      -  \frac{32 \pi r^{2}}{3(1-f)} \tilde{T}_{tt}
      ,
      \label{eq:-Ein-non-vac-XF-constraint-apha-Phie-with-SLambdaF-l=1}
  \\
  f Y_{(e)}
  &=&
      r f \partial_{t}\left(
      X_{(e)}
      + \frac{r}{2} \partial_{r}\tilde{F}
      \right)
      +  \frac{3f-1}{4} r \partial_{t}\tilde{F}
      + 8 \pi r^{2} f \tilde{T}_{tr}
      \nonumber\\
  &=&
      (1-f) \partial_{t}\Phi_{(e)}
      -  2 r f \partial_{t}\partial_{r}\Phi_{(e)}
      -  \frac{16 \pi r^{3}}{3(1-f)} \partial_{t}\tilde{T}_{tt}
      + 8 \pi r^{2} f \tilde{T}_{tr}
      ,
      \label{eq:ll+1fYe-1st-pert-Ein-non-vac-tr-FAB-div-t-r-sum-l=1}
\end{eqnarray}
where we used Eq.~(\ref{eq:Moncrief-master-variable-final-sum-fX=l=1})
and (\ref{eq:-Ein-non-vac-XF-constraint-apha-Phie-with-SLambdaF-l=1})
in the derivation of
Eq.~(\ref{eq:ll+1fYe-1st-pert-Ein-non-vac-tr-FAB-div-t-r-sum-l=1}).
Under the given the components $\tilde{T}_{tt}$ and $\tilde{T}_{tr}$
of the linearized energy-momentum tensor,
Eqs.~(\ref{eq:-Ein-non-vac-XF-constraint-apha-Phie-with-SLambdaF-l=1})
and
(\ref{eq:ll+1fYe-1st-pert-Ein-non-vac-tr-FAB-div-t-r-sum-l=1}) yield
that the component $\tilde{F}$ and $Y_{(e)}$ are determined by
$\Phi_{(e)}$.
Furthermore, substituting
Eq.~(\ref{eq:-Ein-non-vac-XF-constraint-apha-Phie-with-SLambdaF-l=1})
into Eq.~(\ref{eq:Moncrief-master-variable-final-sum-fX=l=1}), we obtain
\begin{eqnarray}
  f X_{(e)}
  &=&
      -  \frac{f(1-f)}{r} \Phi_{(e)}
      -  f (1-f) \partial_{r}\Phi_{(e)}
      + 2 r f \partial_{r}( f \partial_{r}\Phi_{(e)} )
      \nonumber\\
  &&
     - 8 \pi r^{2} \tilde{T}_{tt}
     + \frac{16\pi r^{2} f}{(1-f)} \tilde{T}_{tt}
     + \frac{16 \pi r^{3} f}{3(1-f)} \partial_{r}\tilde{T}_{tt}
     .
  \label{eq:Moncrief-master-variable-final-sum-fX=l=1-2}
\end{eqnarray}
This also yields that the component $X_{(e)}$ is determined by
$\Phi_{(e)}$ under the given components of the linearized
energy-momentum tensor.
Thus, the components $X_{(e)}$, $Y_{(e)}$, and $\tilde{F}$ are
determined by the single variable $\Phi_{(e)}$ apart from the
contribution from the components of the linearized energy-momentum
tensor.


The determination of the Moncrief variable $\Phi_{(e)}$ is
accomplished by solving the master equation
(\ref{eq:Zerilli-Moncrief-eq-final-sum}):
\begin{eqnarray}
  -  \frac{1}{f} \partial_{t}^{2}\Phi_{(e)}
  + \partial_{r}\left[ f \partial_{r}\Phi_{(e)} \right]
  -
  \frac{1-f}{r^{2}} \Phi_{(e)}
  =
  16 \pi \frac{r}{\Lambda} S_{(\Phi_{(e)})}
  ,
  \label{eq:Zerilli-Moncrief-eq-final-sum-l=1}
\end{eqnarray}
and the source term in Eq.~(\ref{eq:Zerilli-Moncrief-eq-final-sum}) is
given by
\begin{eqnarray}
  S_{(\Phi_{(e)})}
  &=&
      \frac{1}{2} r \partial_{t}\tilde{T}_{tr}
      -  \frac{1}{2} r \partial_{r}\tilde{T}_{tt}
      + \frac{1-4f}{2f} \tilde{T}_{tt}
      -  \frac{f}{2} \tilde{T}_{rr}
      - f \tilde{T}_{(e1)r}
      .
      \label{eq:SPhie-def-explicit-sum-l=1}
\end{eqnarray}
The master variable $\Phi_{(e)}$ is determined through the master
equation (\ref{eq:Zerilli-Moncrief-eq-final-sum-l=1}) with appropriate
initial conditions.


Furthermore, we have to take into account of the perturbation of the
divergence of the energy-momentum tensor, which are summarized as
follows:
\begin{eqnarray}
  &&
     -  \partial_{t}\tilde{T}_{tt}
     + f^{2} \partial_{r}\tilde{T}_{rt}
     + \frac{(1+f)f}{r} \tilde{T}_{rt}
     -  \frac{2f}{r} \tilde{T}_{(e1)t}
     =
     0
     ,
     \label{eq:div-barTab-linear-AB-t-comp-reduce-l=1}
  \\
  &&
     - \partial_{t}\tilde{T}_{tr}
     + \frac{1-f}{2rf} \tilde{T}_{tt}
     + f^{2} \partial_{r}\tilde{T}_{rr}
     + \frac{(3+f)f}{2r} \tilde{T}_{rr}
     -  \frac{2f}{r} \tilde{T}_{(e1)r}
     -  \frac{f}{r} \tilde{T}_{(e0)}
     =
     0
     ,
     \label{eq:div-barTab-linear-AB-r-comp-reduce-l=1}
     \\
  &&
     - \partial_{t}\tilde{T}_{(e1)t}
     + f^{2} \partial_{r}\tilde{T}_{(e1)r}
     + \frac{(1+2f)f}{r} \tilde{T}_{(e1)r}
     + \frac{f}{2r} \tilde{T}_{(e0)}
     =
     0
     .
     \label{eq:div-barTab-linear-p-mode-reduce-l=1}
\end{eqnarray}
The expression of (\ref{eq:SPhie-def-explicit-sum-l=1}) for the source
term $S_{(\Phi_{(e)})}$ in Eq.~(\ref{eq:SPhie-def-explicit-sum-l=1})
was derived by using
Eq.~(\ref{eq:div-barTab-linear-AB-r-comp-reduce-l=1}).


\subsection{$l=1$ mode vacuum case}
\label{sec:l=1_Schwarzschild_Background-vac}


As in the case of $l=0$ modes, it is instructive to consider the
vacuum case where all components of the linearized energy-momentum
tensor vanish before the derivation of the non-vacuum case.


Here, we consider the covariant form $\ScrF_{ab}$ of the $l=1$-mode
metric perturbation as follows:
\begin{eqnarray}
  \ScrF_{ab}
  =
  \sum_{m=-1}^{1} \tilde{F}_{AB} k_{(\Delta+2)m} (dx^{A})_{a}(dx^{B})_{b}
  +
  \frac{1}{2} \sum_{m=-1}^{1} \gamma_{pq} r^{2} \tilde{F} k_{(\Delta+2)m} (dx^{p})_{a} (dx^{q})_{b}
  .
  \label{eq:2+2-gauge-inv-var-even-cov-mode-l=1}
\end{eqnarray}
The harmonic function $k_{(\Delta+2)m}$ is explicitly given by
Eqs.~(\ref{eq:l=1-m=0-mode-func-explicit}) and
(\ref{eq:l=1-m=pm1-mode-func-explicit}).
If we impose the regularity on these harmonics by the choice
$\delta=0$, these harmonics are given by the spherical harmonics $Y_{l=1,m}$
with $l=1$:
\begin{eqnarray}
  \label{eq:Ylm-l=1}
  Y_{l=1,m=0} \propto \cos\theta, \quad Y_{l=1,m=1} \propto \sin\theta e^{i\phi},
  \quad Y_{l=1,m=-1} \propto \sin\theta e^{-i\phi}.
\end{eqnarray}
Since the extension of our arguments to $m=\pm 1$ modes is
straightforward, we concentrate only on the $m=0$ modes.


For the $m=0$ mode, the gauge-invariant part $\ScrF_{ab}$ of the
metric perturbation is given by
\begin{eqnarray}
  \ScrF_{ab}
  &=&
      \left(
      f X_{(e)}
      \right) \cos\theta \left\{
      - (dt)_{a} (dt)_{b} - f^{-2} (dr)_{a} (dr)_{b}
      \right\}
      + \frac{2}{f} (f Y_{(e)}) \cos\theta (dt)_{(a} (dr)_{b)}
      \nonumber\\
  &&
      +
      \frac{1}{2} r^{2} \tilde{F} \cos\theta \left\{
      (d\theta)_{a} (d\theta)_{b}
      +
      \sin^{2}\theta (d\phi)_{a} (d\phi)_{b}
      \right\}
      .
      \label{eq:2+2-gauge-inv-var-even-cov-mode-l=1-m=0}
\end{eqnarray}
By choosing $\tilde{T}_{tt}=\tilde{T}_{tr}=0$ in
Eqs.~(\ref{eq:-Ein-non-vac-XF-constraint-apha-Phie-with-SLambdaF-l=1}),
(\ref{eq:ll+1fYe-1st-pert-Ein-non-vac-tr-FAB-div-t-r-sum-l=1}), and
(\ref{eq:Moncrief-master-variable-final-sum-fX=l=1-2}),
we obtain the vacuum solutions $\tilde{F}$, $Y_{(e)}$, and $X_{(e)}$
of the metric perturbation as follows:
\begin{eqnarray}
  X_{(e)}
  &=&
      -  \frac{1}{r} (1-f) \Phi_{(e)}
      -  (1-f) \partial_{r}\Phi_{(e)}
      + 2 r \partial_{r}\left[ f \partial_{r}\Phi_{(e)} \right]
      ,
      \label{eq:fXe-sol-l=1-vac-m=0}
  \\
  Y_{(e)}
  &=&
      \partial_{t}\left[
      + \frac{1}{f} (1-f) \Phi_{(e)}
      -  2 r \partial_{r}\Phi_{(e)}
      \right]
      ,
      \label{eq:Ye-solution-l=1-vac-m=0}
  \\
  \tilde{F}
  &=&
      -  4 f \partial_{r}\Phi_{(e)}
      -  4 \frac{1-f}{r} \Phi_{(e)}
      .
      \label{eq:-Ein-non-vac-XF-constraint-Phie-reduce-l=1-vac-m=0}
\end{eqnarray}
Here, $\Phi_{(e)}$ is a solution to the equation
\begin{eqnarray}
  -  \frac{1}{f} \partial_{t}^{2}\Phi_{(e)}
  + \partial_{r}\left[ f \partial_{r}\Phi_{(e)} \right]
  -
  \frac{1-f}{r^{2}} \Phi_{(e)}
  =
  0
  .
  \label{eq:Zerilli-Moncrief-eq-final-sum-l=1-vac}
\end{eqnarray}


As in the case of $l=0$ mode, we consider the problem whether the
solution (\ref{eq:2+2-gauge-inv-var-even-cov-mode-l=1-m=0}) with
Eqs.~(\ref{eq:fXe-sol-l=1-vac-m=0})--(\ref{eq:-Ein-non-vac-XF-constraint-Phie-reduce-l=1-vac-m=0})
is described by ${\pounds}_{V}g_{ab}$ for an appropriate vector field
$V_{a}$, or not.
From the symmetry of the above solution, we consider the case where
the vector field $V_{a}$ is given by
\begin{eqnarray}
  \label{eq:Vphi-0-l=1-m=0}
  V_{a} = V_{t} (dt)_{a} + V_{r} (dr)_{a} + V_{\theta} (d\theta)_{a},
  \quad
  \partial_{\phi}V_{t} =\partial_{\phi}V_{r}
  = \partial_{\phi}V_{\theta} = 0
\end{eqnarray}
and calculate all components of ${\pounds}_{V}g_{ab}$.
We note that all components of $\ScrF_{ab}$ given by
Eq.~(\ref{eq:2+2-gauge-inv-var-even-cov-mode-l=1-m=0}) are
proportional to $\cos\theta$.
Therefore, if we may identify some components of $\ScrF_{ab}$ with
${\pounds}_{V}g_{ab}$, the $\theta$-dependence of the components in
Eq.~(\ref{eq:Vphi-0-l=1-m=0}) should be given by
\begin{eqnarray}
  \label{eq:Vphi-0-l=1-m=0-theta-dependence}
  V_{a} = v_{t}(t,r) \cos\theta (dt)_{a} + v_{r} \cos\theta (dr)_{a} +
  v_{\theta} \sin\theta (d\theta)_{a}.
\end{eqnarray}
Then, the non-trivial components of ${\pounds}_{V}g_{ab}$ are given by
\begin{eqnarray}
  \label{eq:poundsVgab-components-tt-l=1-m=0}
  &&
     {\pounds}_{V}g_{tt}
     =
     \left(2 \partial_{t}v_{t} - f f' v_{r}\right) \cos\theta
     \neq
     0
     ,
  \\
  \label{eq:poundsVgab-components-tr-l=1-m=0}
  &&
     {\pounds}_{V}g_{tr}
     =
     \left(
     \partial_{t}v_{r} + \partial_{r}v_{t} - \frac{f'}{f} v_{t}
     \right) \cos\theta
     \neq
     0
     ,
  \\
  \label{eq:poundsVgab-components-ttheta-l=1-m=0}
  &&
     {\pounds}_{V}g_{t\theta}
     =
     \left(\partial_{t}v_{\theta} - v_{t}\right) \sin\theta
     =
     0
     ,
  \\
  \label{eq:poundsVgab-components-rr-l=1-m=0}
  &&
     {\pounds}_{V}g_{rr}
     =
     2 f^{-1/2} \partial_{r}\left(
     f^{1/2} v_{r}
     \right) \cos\theta
     \neq
     0
     ,
  \\
  \label{eq:poundsVgab-components-rtheta-l=1-m=0}
  &&
     {\pounds}_{V}g_{r\theta}
     =
     \left(
     r^{2} \partial_{r}\left(\frac{1}{r^{2}} v_{\theta} \right) - v_{r}
     \right) \sin\theta
     =
     0
     ,
  \\
  \label{eq:poundsVgab-components-thetatheta-l=1-m=0}
  &&
     {\pounds}_{V}g_{\theta\theta}
     =
     2 \left(
     v_{\theta} + rf v_{r}
     \right)
     \cos\theta
     \neq
     0
     ,
  \\
  \label{eq:poundsVgab-components-phiphi-l=1-m=0}
  &&
     {\pounds}_{V}g_{\phi\phi}
     =
     2 \left(
     rf v_{r} + v_{\theta}
     \right)
     \sin^{2}\theta \cos\theta
     \neq
     0
     .
\end{eqnarray}
From Eqs.~(\ref{eq:poundsVgab-components-ttheta-l=1-m=0}) and
(\ref{eq:poundsVgab-components-rtheta-l=1-m=0}), we obtain
\begin{eqnarray}
  r^{2}v(t,r) := v_{\theta}, \quad
  v_{t} = \partial_{t}v_{\theta} = r^{2}\partial_{t}v, \quad
  v_{r} = r^{2}\partial_{r}\left(\frac{1}{r^{2}}v_{\theta}\right) =
  r^{2}\partial_{r}v,
\end{eqnarray}
i.e.,
\begin{eqnarray}
  \label{eq:Vphi-0-l=1-m=0-theta-dependence-2}
  V_{a}
  =
  r^{2}\partial_{t}v \cos\theta (dt)_{a}
  + r^{2}\partial_{r}v \cos\theta (dr)_{a}
  + r^{2} v \sin\theta (d\theta)_{a}.
\end{eqnarray}
Then,
Eqs.~(\ref{eq:poundsVgab-components-tt-l=1-m=0})--(\ref{eq:poundsVgab-components-phiphi-l=1-m=0})
are summarized as
\begin{eqnarray}
  \label{eq:poundsVgab-components-tt-l=1-m=0-2}
  &&
     {\pounds}_{V}g_{tt}
     =
     r^{2}\left(2 \partial_{t}^{2}v - \frac{f(1-f)}{r} \partial_{r}v\right) \cos\theta
     ,
  \\
  \label{eq:poundsVgab-components-tr-l=1-m=0-2}
  &&
     {\pounds}_{V}g_{tr}
     =
     \partial_{t}\left(
     2 r^{2} \partial_{r}v
     -  \frac{1-3f}{f} r v
     \right) \cos\theta
     ,
  \\
  \label{eq:poundsVgab-components-rr-l=1-m=0-2}
  &&
     {\pounds}_{V}g_{rr}
     =
     2 f^{-1/2} \partial_{r}\left( f^{1/2} r^{2}\partial_{r}v \right) \cos\theta
     ,
  \\
  \label{eq:poundsVgab-components-thetatheta-l=1-m=0-2}
  &&
     {\pounds}_{V}g_{\theta\theta}
     =
     2 r^{2}\left(
     rf \partial_{r}v + v
     \right)
     \cos\theta
     .
\end{eqnarray}


As the first trial, we consider the correspondence
\begin{eqnarray}
  {\pounds}_{V}g_{\theta\theta} = \ScrF_{\theta\theta},
\end{eqnarray}
i.e.,
\begin{eqnarray}
  rf \partial_{r}v
  + v
  =
  -  f \partial_{r}\Phi_{(e)}
  -  \frac{1-f}{r} \Phi_{(e)}
  .
  \label{eq:poundsVgthetatheta-trial}
\end{eqnarray}
As the second trial, we consider the correspondence
\begin{eqnarray}
  {\pounds}_{V}g_{rr} = \ScrF_{rr},
\end{eqnarray}
i.e.,
\begin{eqnarray}
  -  \frac{1-5f}{f} r \partial_{r}v
  + \frac{2r^{2}}{f} \partial_{r}\left( f \partial_{r}v \right)
  =
  \frac{1-f}{rf} \Phi_{(e)}
  + \frac{1-f}{f} \partial_{r}\Phi_{(e)}
  -  \frac{2r}{f} \partial_{r}\left[ f \partial_{r}\Phi_{(e)} \right]
  .
  \label{eq:poundsVgrr-trial}
\end{eqnarray}
From Eqs.~(\ref{eq:poundsVgthetatheta-trial}) and
(\ref{eq:poundsVgrr-trial}), we obtain
\begin{eqnarray}
  v
  =
  - \frac{1}{r} \Phi_{(e)}
  .
  \label{eq:poundsVgrr-poundsVgthetatheta-trial-result}
\end{eqnarray}
Substituting Eq.~(\ref{eq:poundsVgrr-poundsVgthetatheta-trial-result})
into Eq.~(\ref{eq:poundsVgab-components-tr-l=1-m=0}), we obtain
\begin{eqnarray}
  {\pounds}_{V}g_{tr}
  =
  \partial_{t}\left(
  -  2 r \partial_{r}\Phi_{(e)}
  + \frac{1}{f} (1-f) \Phi_{(e)}
  \right) \cos\theta
  =
  \ScrF_{tr}
  .
  \label{eq:poundsVgtr-trial-result}
\end{eqnarray}
Furthermore, substituting
Eq.~(\ref{eq:poundsVgrr-poundsVgthetatheta-trial-result}) into
Eq.~(\ref{eq:poundsVgab-components-tt-l=1-m=0}), we obtain
\begin{eqnarray}
  {\pounds}_{V}g_{tt}
  &=&
      \left(
      - 2 r \partial_{t}^{2}\Phi_{(e)}
      - \frac{f(1-f)}{r} \Phi_{(e)}
      + f(1-f) \partial_{r}\Phi_{(e)}
      \right) \cos\theta
      \nonumber\\
  &=&
      - f \left(
      - \frac{1}{r} (1-f) \Phi_{(e)}
      - (1-f) \partial_{r}\Phi_{(e)}
      + 2 r \partial_{r}\left[ f \partial_{r}\Phi_{(e)} \right]
      \right) \cos\theta
      \nonumber\\
  &=&
      \ScrF_{tt}
      ,
      \label{eq:poundsVgtt-trial-result}
\end{eqnarray}
where we used Eq.~(\ref{eq:Zerilli-Moncrief-eq-final-sum-l=1-vac}).


Then, we have shown that
\begin{eqnarray}
  \ScrF_{ab}
  =
  {\pounds}_{V}g_{ab}
  ,
  \label{eq:2+2-gauge-inv-var-even-cov-mode-l=1-m=0-tr-24}
\end{eqnarray}
where
\begin{eqnarray}
  V_{a}
  =
  - r \partial_{t}\Phi_{(e)} \cos\theta (dt)_{a}
  +
  \left( \Phi_{(e)} - r \partial_{r}\Phi_{(e)} \right) \cos\theta (dr)_{a}
  - r \Phi_{(e)} \sin\theta (d\theta)_{a}
  .
  \label{eq:generator-covariant-vacuum-l=1-m=0-result}
\end{eqnarray}
Thus, the vacuum solution of $l=1$-mode perturbations described by the
Lie derivative of the background metric through the master equation
(\ref{eq:Zerilli-Moncrief-eq-final-sum-l=1-vac}).


\subsection{$l=1$ mode non-vacuum case}
\label{sec:even_l=1_Schwarzschild_Background-non-vac}


Here, we consider the non-vacuum solution to the $l=1$ even-mode
linearized Einstein equations.
In this non-vacuum case, we concentrate only on the $m=0$ mode
perturbations as in the vacuum case, because the extension to our
arguments to $m=\pm 1$ modes is straightforward.
The solution is given by the covariant form
(\ref{eq:2+2-gauge-inv-var-even-cov-mode-l=1-m=0}) as in the case of
the vacuum case.
The non-vacuum solutions for the variable $\tilde{F}$, $Y_{(e)}$, and
$X_{(e)}$ are given by
Eqs.~(\ref{eq:-Ein-non-vac-XF-constraint-apha-Phie-with-SLambdaF-l=1}),
(\ref{eq:ll+1fYe-1st-pert-Ein-non-vac-tr-FAB-div-t-r-sum-l=1}), and
(\ref{eq:Moncrief-master-variable-final-sum-fX=l=1-2}), respectively.
The master variable $\Phi_{(e)}$ must satisfy the master equation
(\ref{eq:Zerilli-Moncrief-eq-final-sum-l=1}) with the source term
(\ref{eq:SPhie-def-explicit-sum-l=1}).
We have to emphasize that the components of the linear perturbation of
energy-momentum tensor satisfy the continuity equations (\ref{eq:div-barTab-linear-AB-t-comp-reduce-l=1})--(\ref{eq:div-barTab-linear-p-mode-reduce-l=1}).
Then, the components of the gauge-invariant part $\ScrF_{ab}$ for
$l=1$ even-mode non-vacuum perturbations are summarized as follows:
\begin{eqnarray}
  \ScrF_{tt}
  &=&
      f \left[
      \frac{1}{r} (1-f) \Phi_{(e)}
      + (1-f) \partial_{r}\Phi_{(e)}
      -  2 r \partial_{r}\left[ f \partial_{r}\Phi_{(e)} \right]
      \right]  \cos\theta
      \nonumber\\
  &&
     + \frac{8 \pi r^{2}}{3(1-f)} \left[
     3(1-3f) \tilde{T}_{tt}
     -  2 r f \partial_{r}\tilde{T}_{tt}
     \right]  \cos\theta
     ,
     \label{eq:tildeFtt-l=1-m=0-nonvac-sum}
  \\
  \ScrF_{tr}
  &=&
      r \partial_{t}\left[
      \frac{1-f}{rf} \Phi_{(e)}
      -  2 \partial_{r}\Phi_{(e)}
      -  \frac{16 \pi r^{2}}{3f(1-f)} \tilde{T}_{tt}
      \right] \cos\theta
      + 8 \pi r^{2} \tilde{T}_{tr} \cos\theta
      ,
     \label{eq:tildeFtr-l=1-m=0-nonvacsum}
  \\
  \ScrF_{rr}
  &=&
      \frac{1}{f} \left[
      \frac{1-f}{r} \Phi_{(e)}
      + (1-f) \partial_{r}\Phi_{(e)}
      -  2 r \partial_{r}( f \partial_{r}\Phi_{(e)} )
      \right] \cos\theta
      \nonumber\\
  &&
      + \frac{8 \pi r^{2}}{3f^{2}(1-f)}
      \left[
      3(1-3f) \tilde{T}_{tt}
      -  2 r f \partial_{r}\tilde{T}_{tt}
      \right] \cos\theta
      ,
      \label{eq:tildeFrr-l=1-m=0-nonvacsum}
  \\
  \ScrF_{\theta\theta}
  &=&
      - 2 r \left[
      r f \partial_{r}\Phi_{(e)}
      + (1-f) \Phi_{(e)}
      + \frac{8 \pi r^{3}}{3(1-f)} \tilde{T}_{tt}
      \right] \cos\theta
      ,
      \label{eq:tildeFthetatheta-l=1-m=0-nonvacsum}
  \\
  \ScrF_{\phi\phi}
  &=&
      - 2 r \left[
      r f \partial_{r}\Phi_{(e)}
      + (1-f) \Phi_{(e)}
      + \frac{8 \pi r^{3}}{3(1-f)} \tilde{T}_{tt}
      \right] \sin^{2}\theta \cos\theta
      .
      \label{eq:tildeFtphiphi-l=1-m=0-nonvacsum}
\end{eqnarray}


As seen in the vacuum case, if we choose the generator $V_{a}$ as
Eq.~(\ref{eq:generator-covariant-vacuum-l=1-m=0-result}), i.e.,
\begin{eqnarray}
  V_{a} = V_{(vac)a}
  &:=&
       -  r \partial_{t}\Phi_{(e)} \cos\theta (dt)_{a}
       + \left( \Phi_{(e)} - r \partial_{r}\Phi_{(e)} \right) \cos\theta (dr)_{a}
       \nonumber\\
  &&
     -  r \Phi_{(e)} \sin\theta (d\theta)_{a}
     ,
     \label{eq:generator-covariant-vacuum-l=1-m=0-result-2}
\end{eqnarray}
we obtain
\begin{eqnarray}
  \label{eq:poundsVgtt-l=1-m=0-vac}
  &&
     {\pounds}_{V}g_{tt}
     =
     f \left[
     - \frac{2r}{f} \partial_{t}^{2}\Phi_{(e)}
     + (1-f) \partial_{r}\Phi_{(e)}
     -  \frac{1-f}{r} \Phi_{(e)}
     \right] \cos\theta
     ,
  \\
  \label{eq:poundsVgtr-l=1-m=0-vac}
  &&
     {\pounds}_{V}g_{tr}
     =
     r \partial_{t}\left(
     \frac{1-f}{rf} \Phi_{(e)}
     -  2 \partial_{r}\Phi_{(e)}
     \right) \cos\theta
     ,
  \\
  \label{eq:poundsVgrr-l=1-m=0-vac}
  &&
     {\pounds}_{V}g_{rr}
     =
     \frac{1}{f} \left[
     \frac{1-f}{r} \Phi_{(e)}
     + (1-f) \partial_{r}\Phi_{(e)}
     - 2 r \partial_{r}\left( f \partial_{r}\Phi_{(e)} \right)
     \right] \cos\theta
     ,
  \\
  \label{eq:poundsVgthetatheta-l=1-m=0-vac}
  &&
     {\pounds}_{V}g_{\theta\theta}
     =
     - 2 r \left[
     r f \partial_{r}\Phi_{(e)}
     + (1-f) \Phi_{(e)}
     \right]
     \cos\theta
     ,
  \\
  \label{eq:poundsVgphiphi-l=1-m=0-vac}
  &&
     {\pounds}_{V}g_{\phi\phi}
     =
     - 2 r\left(
     r f \partial_{r}\Phi_{(e)}
     + (1-f) \Phi_{(e)}
     \right)
     \sin^{2}\theta\cos\theta
     ,
  \\
  \label{eq:poundsVgttheta-gtphi-grtheta-grphi-with-generator-Va-vac-l=1-m=0}
  &&
     {\pounds}_{V}g_{t\theta}
     =
     {\pounds}_{V}g_{t\phi} =
     {\pounds}_{V}g_{r\theta} = {\pounds}_{V}g_{r\phi} =
     {\pounds}_{V}g_{\theta\phi} = 0
     .
\end{eqnarray}
Through these formulae of the components ${\pounds}_{V}g_{ab}$
and
Eqs.~(\ref{eq:tildeFtt-l=1-m=0-nonvac-sum})--(\ref{eq:tildeFtphiphi-l=1-m=0-nonvacsum})
for the components of $\ScrF_{ab}$, we obtain
\begin{eqnarray}
  \ScrF_{tt}
  &=&
      {\pounds}_{V}g_{tt}
      - \frac{16 \pi r^{2} f^{2}}{3(1-f)} \left[
      \frac{1+f}{2} \tilde{T}_{rr}
      + r f \partial_{r}\tilde{T}_{rr}
      -  \tilde{T}_{(e0)}
      -  4 \tilde{T}_{(e1)r}
      \right]  \cos\theta
      ,
     \label{eq:tildeFtt-l=1-m=0-nonvac-sum-2}
  \\
  \ScrF_{tr}
  &=&
      {\pounds}_{V}g_{tr}
      - \frac{16 \pi r^{3}}{3f(1-f)} \left[
      \partial_{t}\tilde{T}_{tt}
      - \frac{3f(1-f)}{2r} \tilde{T}_{tr}
      \right] \cos\theta
      ,
     \label{eq:tildeFtr-l=1-m=0-nonvacsum-2}
  \\
  \ScrF_{rr}
  &=&
      {\pounds}_{V}g_{rr}
      - \frac{16 \pi r^{3}}{3f(1-f)}
      \left[
      \partial_{r}\tilde{T}_{tt}
      - \frac{3(1-3f)}{2rf} \tilde{T}_{tt}
      \right] \cos\theta
      ,
      \label{eq:tildeFrr-l=1-m=0-nonvacsum-2}
  \\
  \ScrF_{\theta\theta}
  &=&
      {\pounds}_{V}g_{\theta\theta}
      - \frac{16 \pi r^{4}}{3(1-f)} \tilde{T}_{tt} \cos\theta
      ,
      \label{eq:tildeFthetatheta-l=1-m=0-nonvacsum-2}
  \\
  \ScrF_{\phi\phi}
  &=&
      {\pounds}_{V}g_{\phi\phi}
      - \frac{16 \pi r^{4}}{3(1-f)} \tilde{T}_{tt} \sin^{2}\theta \cos\theta
      ,
      \label{eq:tildeFphiphi-l=1-m=0-nonvacsum-2}
\end{eqnarray}
where we used Eq.~(\ref{eq:Zerilli-Moncrief-eq-final-sum-l=1}) with
the source term (\ref{eq:SPhie-def-explicit-sum-l=1}) and the
component (\ref{eq:div-barTab-linear-AB-r-comp-reduce-l=1}) of the
continuity equation in Eq.~(\ref{eq:tildeFtt-l=1-m=0-nonvac-sum-2}).
Equations~(\ref{eq:tildeFtt-l=1-m=0-nonvac-sum-2})--(\ref{eq:tildeFphiphi-l=1-m=0-nonvacsum-2})
are summarized as
\begin{eqnarray}
  \ScrF_{ab}
  &=&
      {\pounds}_{V}g_{ab}
      - \frac{16 \pi r^{2}}{3(1-f)}\left[
      f^{2} \left\{
      \frac{1+f}{2} \tilde{T}_{rr}
      + r f \partial_{r}\tilde{T}_{rr}
      -  \tilde{T}_{(e0)}
      -  4 \tilde{T}_{(e1)r}
      \right\}  (dt)_{a}(dt)_{b}
      \right.
      \nonumber\\
  && \quad\quad\quad\quad\quad\quad\quad\quad\quad\quad
     \left.
      + \frac{2r}{f} \left\{
      \partial_{t}\tilde{T}_{tt}
      - \frac{3f(1-f)}{2r} \tilde{T}_{tr}
      \right\} (dt)_{(a}(dr)_{b)}
     \right.
      \nonumber\\
  && \quad\quad\quad\quad\quad\quad\quad\quad\quad\quad
     \left.
      + \frac{r}{f}
      \left\{
      \partial_{r}\tilde{T}_{tt}
      - \frac{3(1-3f)}{2rf} \tilde{T}_{tt}
      \right\} (dr)_{a}(dr)_{b}
     \right.
      \nonumber\\
  && \quad\quad\quad\quad\quad\quad\quad\quad\quad\quad
     \left.
      + r^{2} \tilde{T}_{tt} \gamma_{ab}
     \right] \cos\theta
      .
      \label{eq:tildeFab-l=1-m=0-nonvacsum-2-cov}
\end{eqnarray}
We note that there may be exist the term ${\pounds}_{W}g_{ab}$ in the
right-hand side of Eqs.~(\ref{eq:tildeFab-l=1-m=0-nonvacsum-2-cov}) in
addition to the term ${\pounds}_{V}g_{ab}$ discussed above.
Such term will depend on the equation of state of the matter field.
This situation can be seen in the Part III
paper~\cite{K.Nakamura-2021e}.
Even if we consider such terms, we will not have a simple
expression of the metric perturbation, in general.
Therefore, we will not carry out such further considerations, here.


\section{Summary and Discussion}
\label{sec:Summary_and_Discussion}


In summary, after reviewing our general framework of the
general-relativistic gauge-invariant perturbation theory and our
strategy for the linear perturbations on the Schwarzschild background
spacetime proposed in Refs.~\cite{K.Nakamura-2021a,K.Nakamura-2021c},
we developed the component treatments of the even-mode linearized
Einstein equations.
Our proposal in Refs.~\cite{K.Nakamura-2021a,K.Nakamura-2021c} was on
the gauge-invariant treatments of the $l=0,1$ mode perturbations on
the Schwarzschild background spacetime.
Since we used singular harmonic functions at once in our proposal, we
have to confirm whether our proposal is physically reasonable, or not.


To confirm this, in the Part I paper~\cite{K.Nakamura-2021c}, we
carefully discussed the solutions to the Einstein equations for
odd-mode perturbations.
We obtain the Kerr parameter perturbations in the vacuum case, which
is physically reasonable.
In this paper, we carefully discussed the solutions to the even-mode
perturbations.
Due to
Proposal~\ref{proposal:treatment-proposal-on-pert-on-spherical-BG},
we can treat the $l=0,1$ mode perturbations through the equivalent
manner to the $l\geq 2$-mode perturbations.
For this reason, we derive the equations for even-mode perturbations
without making distinction among $l\geq 0$ modes for even-mode
perturbations.


To derive the even-mode perturbations, it is convenient to introduce
the Moncrief variable.
In this paper, we explain the introduction of the Moncrief variable
through an initial value constraint
(\ref{eq:1st-pert-Ein-non-vac-tt-FAB-div-t+r-constraint}) is regard as
an equation for the component $\tilde{F}$ of the metric perturbation
and the Moncrief variable $\Phi_{(e)}$.
This consideration leads to the well-known definition of the Moncrief
variable $\Phi_{(e)}$.
Furthermore, from the evolution equation
(\ref{eq:1st-pert-Ein-non-vac-tt-sum-2-2}), we obtain the well-known
master equation (\ref{eq:Zerilli-Moncrief-eq-final-sum}) for the
Moncrief variable $\Phi_{(e)}$.


Moreover, we obtain the constraint equations
(\ref{eq:-Ein-non-vac-XF-constraint-apha-Phie-with-SLambdaF-sum})
and (\ref{eq:ll+1fYe-1st-pert-Ein-non-vac-tr-FAB-div-t-r-sum-3})
together with the definition
(\ref{eq:Moncrief-master-variable-final-sum-fX=}) of the Moncrief
variable.
From their derivations, we have shown that these equations are valid
not only for $l\geq 2$ but also for $l=0,1$ modes.
We also checked the consistency of these equations, and we derived the
identity of the source terms which are given by the components of the
linear perturbation of the energy-momentum tensor.
This identity is confirmed by the components of the linear
perturbation of the energy-momentum tensor.


In this paper, we also carefully discussed the $l=0,1$ mode solutions
to the linearized Einstein equations for even-mode perturbations to
check that
Proposal~\ref{proposal:treatment-proposal-on-pert-on-spherical-BG} is
physically reasonable.


The $l=0$-mode solutions are discussed in
Sec.~\ref{sec:l=0_Schwarzschild_Background-non-vac}.
After summarizing the linearized Einstein equations and the linearized
continuity equations for generic matter field for $l=0$ mode, we first
considered the vacuum solution of the $l=0$-mode perturbations
following
Proposal~\ref{proposal:treatment-proposal-on-pert-on-spherical-BG}.
Then, we showed that the additional mass parameter perturbation of the
Schwarzschild spacetime is the only solution apart from the terms of
the Lie derivative of the background metric $g_{ab}$ in the vacuum
case.
This is the gauge-invariant realization of the linearized version of
the Birkhoff theorem~\cite{S.W.Hawking-G.F.R.Ellis-1973}.


In the non-vacuum case, we use the method of the variational constant
with the Schwarzschild mass constant parameter in vacuum case.
Then, we obtained the general non-vacuum solution to the linearized
Einstein equation for the $l=0$ mode.
As the result, we obtained the linearized metric
(\ref{eq:calFab+poundsVg-l=0-non-vac-final}).
The solution (\ref{eq:calFab+poundsVg-l=0-non-vac-final}) includes the
additional mass parameter perturbation $M_{1}$ of the Schwarzschild
mass and the integration of the energy density.
Furthermore, in the solution
(\ref{eq:calFab+poundsVg-l=0-non-vac-final}),  we have the
$2(dt)_{(a}(dr)_{b)}$ term due to the integration of the components of
the energy-momentum tensor.
In the solution (\ref{eq:calFab+poundsVg-l=0-non-vac-final}), we also
have the term which have the form of the Lie derivative of the
background metric $g_{ab}$.
The off-diagonal term of $2(dt)_{(a}(dr)_{b)}$ can be eliminate by the
replacement of the generator $V_{a}$ of the term of the Lie derivative
of the $g_{ab}$.
However, if we eliminate the off-diagonal term of
$2(dt)_{(a}(dr)_{b)}$ through the replacement of the generator
$V_{a}$, we have additional term to the diagonal components of the
linearized metric perturbation
(\ref{eq:calFab+poundsVg-l=0-non-vac-final}).
Since these diagonal components have complicated forms, we do not
carry out this displacement.


We also discussed the $l=1$-mode perturbations in
Sec.~\ref{sec:l=1_Schwarzschild_Background-non-vac}.
In this paper, we concentrated only on the $m=0$ mode, since the
extension to $m=\pm 1$ modes are straightforward.
The solution of the $l=1$ mode is obtained through the similar
strategy to the case of $l\geq 2$ modes that are discussed in
Sec.~\ref{sec:Schwarzschild_Background-non-vaccum-even-treatment}.
As in the case of $l=0$-mode perturbations, we first discuss the
vacuum solution for $l=1$-mode perturbations.
As the result, $l=1$-mode vacuum metric perturbations are described by
the Lie derivative of the background metric $g_{ab}$ with an
appropriate operator.
On the other hand, in the non-vacuum $l=1$-mode perturbations, the
$l=1$ mode metric perturbation have the contribution from the
components of the energy-momentum tensor of the matter field in
addition to the term of the Lie derivative of the background metric
$g_{ab}$ which is derived as the above vacuum solution.


As the odd-mode solutions in the Part I paper~\cite{K.Nakamura-2021c},
we also have the terms of the Lie derivative of the background metric
$g_{ab}$ in the derived solutions in the $l=0,1$ even-mode solutions.
We have to remind that our definition of gauge-invariant variables is
not unique, and we may always add the term of the Lie derivative of
the background metric $g_{ab}$ with a gauge-invariant generator as
emphasized in
Sec.~\ref{sec:general-framework-GI-perturbation-theroy}.
In other words, we may have such terms in derived solutions at any
time, and we may say that the appearance of such terms is a natural
consequence due to the symmetry in the definition of gauge-invariant
variables.
Furthermore, since our formulation completely excludes the second kind
gauge through
Proposal~\ref{proposal:treatment-proposal-on-pert-on-spherical-BG},
these terms of the Lie derivative should be regarded as the degree of
freedom of the first kind gauge, i.e., the coordinate transformation
of the physical spacetime $\ScrM_{\epsilon}$ as emphasized in the
Part I paper~\cite{K.Nakamura-2021c}.
This discussion is the consequence of our distinction of the first-
and second-kind of gauges and the complete exclusion of the gauge
degree of freedom of the second kind as emphasized in the Part I
paper~\cite{K.Nakamura-2021c}.


We also note that the existence of the additional mass parameter
perturbation $M_{1}$ requires the perturbations of $\tilde{F}$ due to
the linearized Einstein equations.
In this sense, the term described by the Lie derivative of the
background spacetime is necessary.
The solutions derived in this paper and the Part I
paper~\cite{K.Nakamura-2021c} are local perturbative solutions.
If we construct the global solution, we have to use the solutions
obtained in this paper and in the Part I paper~\cite{K.Nakamura-2021c}
as local solutions and have to match these local solutions.
We expect that the term of the Lie derivative derived here will play
important roles in this case.


Besides the term of the Lie derivative of the background metric
$g_{ab}$, we have realized the Birkhoff theorem for $l=0$ even-mode
solutions and the Kerr parameter perturbations in $l=1$ odd-mode
solutions.
These solutions are physically reasonable.
This also implies that
Proposal~\ref{proposal:treatment-proposal-on-pert-on-spherical-BG} is
physically reasonable nevertheless we used singular mode functions at
once to construct gauge-invariant variables and imposed the regular
boundary condition on the functions on $S^{2}$ when we solve the
linearized Einstein equations, while the conventional treatment
through the decomposition by the spherical harmonics $Y_{lm}$
corresponds to the imposition of the regular boundary condition from
the starting point.


\section*{Acknowledgements}


The author deeply acknowledged to Professor Hiroyuki Nakano for
various discussions and suggestions.








\end{document}